\documentclass[11pt]{article}
\usepackage[margin=1.15in]{geometry}
\usepackage{amsmath,amssymb,amsthm,mathtools}
\numberwithin{equation}{section}
\usepackage{enumitem}
\usepackage{booktabs}
\usepackage[colorlinks=true,linkcolor=blue,citecolor=blue,urlcolor=blue]{hyperref}
\usepackage{adjustbox}
\usepackage{tikz}
\usetikzlibrary{shapes.geometric,arrows.meta,positioning,calc,fit,backgrounds}
 
\definecolor{fcGreen}{HTML}{E8F5E9}\definecolor{fcGreenE}{HTML}{4CAF50}
\definecolor{fcAmber}{HTML}{FFF3E0}\definecolor{fcAmberE}{HTML}{EF8A00}
\definecolor{fcGrey}{HTML}{F4F4F6}\definecolor{fcGreyE}{HTML}{9E9E9E}
\definecolor{fcViolet}{HTML}{F6F1FC}\definecolor{fcVioletE}{HTML}{7E57C2}
 
\tikzset{
  fcbox/.style   ={draw,rounded corners=2pt,align=center,inner sep=4pt,
                   font=\footnotesize,line width=0.5pt},
  fcstep/.style  ={fcbox,fill=fcGrey,draw=fcGreyE},
  fckey/.style   ={fcbox,fill=fcAmber,draw=fcAmberE},
  fcio/.style    ={fcbox,fill=fcGreen,draw=fcGreenE},
  fcdec/.style   ={draw=fcAmberE,fill=fcAmber,diamond,aspect=2.4,align=center,
                   inner sep=2pt,font=\footnotesize,line width=0.5pt},
  fcflow/.style  ={-{Stealth[length=2mm]},line width=0.5pt,draw=fcGreyE!140},
  fcloop/.style  ={fcflow,draw=fcVioletE},
  fclab/.style   ={font=\scriptsize,inner sep=1.5pt,fill=white},
}
\newcommand{\fcref}[1]{\\[1pt]{\scriptsize #1}}
 
\definecolor{fcBlue}{HTML}{E7F1FB}\definecolor{fcBlueE}{HTML}{3E7CB1}
\definecolor{fcRed}{HTML}{FDEAEA}\definecolor{fcRedE}{HTML}{D9736B}
\definecolor{fcPurple}{HTML}{EDE9F7}
\definecolor{fcGreenBg}{HTML}{F2F9F3}\definecolor{fcAmberBg}{HTML}{FEF8EE}
\tikzset{
  fcin/.style    ={fcbox,fill=fcBlue,draw=fcBlueE},
  fcout/.style   ={fcbox,fill=fcRed,draw=fcRedE},
  fcnote/.style  ={fcbox,fill=fcPurple,draw=fcVioletE,align=left},
  fcplain/.style ={fcbox,align=left},
  fcpanA/.style  ={draw=fcGreenE,dashed,line width=0.6pt,rounded corners=3pt,
                   fill=fcGreenBg},
  fcpanB/.style  ={draw=fcAmberE,dashed,line width=0.6pt,rounded corners=3pt,
                   fill=fcAmberBg},
  fchead/.style  ={font=\footnotesize\bfseries},
  fcline/.style  ={line width=0.5pt,draw=fcGreyE!140},
}

\hypersetup{pdftitle={Hidden parabolic subgroups of GLn over finite chain rings},
pdfauthor={Muhammad Imran}}
\theoremstyle{plain}
\newtheorem{theorem}{Theorem}[section]
\newtheorem{proposition}[theorem]{Proposition}
\newtheorem{lemma}[theorem]{Lemma}
\newtheorem{corollary}[theorem]{Corollary}
\theoremstyle{definition}
\newtheorem{definition}[theorem]{Definition}
\newtheorem{convention}[theorem]{Convention}
\newtheorem{problem}[theorem]{Problem}
\newtheorem{example}[theorem]{Example}
\theoremstyle{remark}
\newtheorem{remark}[theorem]{Remark}

\DeclareMathOperator{\Hom}{Hom}
\DeclareMathOperator{\GL}{GL}
\DeclareMathOperator{\Mat}{M}
\DeclareMathOperator{\tr}{tr}
\DeclareMathOperator{\rank}{rank}
\DeclareMathOperator{\soc}{soc}
\DeclareMathOperator{\Ann}{Ann}

\DeclareMathOperator{\poly}{poly}

\newcommand{\Z}{\mathbb Z}
\newcommand{\Q}{\mathbb Q}
\newcommand{\F}{\mathbb F}
\newcommand{\C}{\mathbb C}
\newcommand{\m}{\mathfrak m}
\newcommand{\p}{\mathfrak p}
\newcommand{\Fl}{\mathcal F}
\newcommand{\hsp}{\mathrm{HSP}}
\newcommand{\rhsp}{\mathrm{rHSP}}
\newcommand{\OK}{\mathcal O_K}
\title{Hidden parabolic subgroups of $\GL_n$ over finite chain rings}
\author{Muhammad Imran
\\
School of Computer Science, University of Birmingham\\
CISITS Lab, Department of Mathematics, Universitas Indonesia\\
e-mail:~\texttt{m.imran.2@bham.ac.uk;~m.imran@sci.ui.ac.id}}
\date{\today}
\begin{document}
\maketitle
\begin{abstract}
We give a bounded-error quantum polynomial-time algorithm for finding hidden parabolic subgroups of $\GL_n(R)$ over any effectively presented finite chain ring $R$.  We call auxiliary group elements or subgroups used to extract information through oracle queries \emph{probes}. For chain length at least two, elementary abelian socle probes first recover the complete flag modulo the maximal ideal. When the residue field has size $q\ge3$, we reconstruct the flag recursively, lifting rank-one terminal members one digit at a time. The intermediate digits reduce to abelian hidden subgroup problems, while the final digit is a finite-field multiple-shift problem solved using learning from subset and Fourier sampling. For $q=2$, where distinct flags can have the same stabilizer, we instead lift the conjugate parabolic subgroup through the congruence filtration. A final linear system determines it modulo normalizer directions. The algorithm outputs generators with probability at least $1-\epsilon$ in time $\poly(n,\log|R|,\log(1/\epsilon))$. Consequences include $\GL_n(\Z/N\Z)$ for every $N\ge2$ and, with an explicit maximal-order presentation, $\GL_n(\OK/N\OK)$ for number fields $K$.
\end{abstract}

\section{Introduction}
The hidden subgroup problem (HSP) asks for generators of a subgroup $H\le G$ from coherent access to a function $f$ constant exactly on its left cosets. 
It is efficiently solvable on quantum computers for finite abelian groups \cite{Kit95,BL95}, but general non-abelian groups remain substantially harder. Efficient algorithms are known for selected group families \cite{ivanyos2001efficient,ivanyos2007efficient,ISS12,Imran-Ivanyos2022,imran2024zero} and selected classes of hidden subgroups. Among the latter, Denney--Moore--Russell and Ivanyos treated Borel stabilizers over finite fields \cite{DMR10,Iva12}, and Decker--Ivanyos--Kulkarni--Qiao--Santha (DIKQS) solved the parabolic HSP in $\GL_n(\F_q)$ for arbitrary parabolic type \cite{DIKQS14}.

We extend the finite field result to finite commutative chain rings $R\supsetneq\m=(\pi)\supsetneq\cdots\supsetneq\m^t=0$ where $\m$ is the maximal ideal of $R$. Write $k=R/\m\cong\F_q$ and $V=R^n$. A parabolic subgroup is the stabilizer of a flag of free direct summands of $V$. Examples of $R$ include $\Z/p^t\Z$, Galois rings, $\F_q[x]/(x^t)$, and prime-power quotients of rings of integers.

\begin{theorem}[Parabolic HSP over finite chain rings]\label{thm:main}
Let $R$ be a finite chain ring in the effective presentation of Section~\ref{sec:model}. Given access to an oracle hiding a parabolic subgroup $H\le\GL_n(R)$ and $0<\epsilon<1$, a quantum algorithm outputs generators of $H$ with probability at least $1-\epsilon$ in time $\poly(n,\log|R|,\log(1/\epsilon))$. The type of the hidden flag is not part of the input.
\end{theorem}

\paragraph{Algorithms.}
For $t=1$ we use DIKQS. Assume $t\ge2$. Our common first step is to recover the reduced flag in $V/\m V\cong k^n$. For a unimodular functional $\mu:V\to R$, the socle group
\[
 S_\mu=\{I+\pi^{t-1}a\mu:a\in V\}\cong(k^n,+)
\]
is elementary abelian. Its intersection with $H$ identifies a member of the reduced flag. Sampling functionals successively in annihilators recovers the entire chain and its type (Proposition~\ref{prop:reduced}).

For $q\ge3$, a parabolic determines its flag (Theorem~\ref{thm:faithful}). If the smallest nonzero member $T$ has rank at least two, an abelian intersection probe produces a nonzero free direct summand $W\subseteq T$, and we recurse on a complement. Rank-one terminal members require a new lifting routine: each non-final digit is an abelian HSP over $k^{n-1}\times k$, whereas the last digit yields the family $F(w-cs)$ with $c\in k^\times$. For small $q$, a Fourier projection produces an instance of learning from subset with support $\{0,1\}$, solved by \cite{IPS18}; for large $q$, direct Fourier sampling suffices. This gives polynomial time uniformly in $q$ and $n$.

For $q=2$, the last rank-one shift has only one copy, and distinct flags can have the same stabilizer. We therefore recover the subgroup directly. After standardizing the reduction, write $H=gPg^{-1}$ with $P$ known and $g\equiv I\pmod\m$. Two-copy socle shifts, each an elementary abelian HSP, strip successive digits of $g$ modulo the block triangular algebra. At the final layer, linear equations determine the remaining conjugator modulo precisely the directions that normalize $P$. This ambiguity does not change $H$.

\paragraph{Why direct transport is insufficient.}
The Frobenius duality and constant-density estimates behind the DIKQS Fourier procedures extend to $R$, but one transported rank-one procedure has repetition factor $(1-1/q)^{-(n-1)}$. There is also a limitation on a different probe family: an algorithm seeing only uniformly sampled pairs $(\lambda,H\cap N_\lambda)$ can require $\Omega(q^{t-1})$ probes to distinguish terminal lines in their last digit. 
This is a restricted transcript lower
bound, not a lower bound for the HSP or for adaptively chosen probes. The new lifting procedures avoid both limitations.

\paragraph{Motivation and organization.}
The pSIDH attack of \cite{CIIKLP23} uses hidden Borel subgroups of $\GL_2(\Z/N\Z)$. As isogeny-based cryptography shifts to higher-dimensional abelian varieties, the problem in higher-dimensional matrix groups $\GL_n(\Z/N\Z)$ might be a useful cryptanalytic tool. Work relating Hawk to a quaternion principal ideal problem \cite{CMEPPW25}, and the directions discussed in \cite{wallet4}, also motivate the case of number-field quotient rings; no reduction of those problems to the parabolic promise is claimed here. Sections~\ref{sec:prelim}--\ref{sec:graded} establish the algebraic setting and reduced-flag algorithm. Sections~\ref{sec:hmslift}--\ref{sec:main} give rank-one lifting and full reconstruction for $q\ge3$; Section~\ref{sec:q2} handles $q=2$. Section~\ref{sec:applications} completes the main proof and gives the arithmetic consequences. Supporting algebra, transported procedures, and arithmetic preprocessing appear in Appendices~\ref{app:algebra}--\ref{app:numberfield}. 

\paragraph{Roadmap of the proof.}
The proof of Theorem~\ref{thm:main} has three stages; the first is common to all $q$.
\begin{enumerate}[label=(\arabic*),leftmargin=*,itemsep=2pt]
\item \emph{Reduced flag} (Section~\ref{sec:graded}). Socle probes $S_\mu$ turn the oracle into elementary abelian HSPs whose answers are members of $\bar\Fl$. Iterating in annihilators recovers $\bar\Fl$ and the type (Proposition~\ref{prop:reduced}).
\item \emph{Case $q\ge3$} (Sections~\ref{sec:hmslift}--\ref{sec:main}). Here $H$ determines $\Fl$ (Theorem~\ref{thm:faithful}), so we reconstruct the flag from the bottom. A nonzero split free $W\subseteq T$ is found either by an abelian intersection probe when $\rank T\ge2$ (Proposition~\ref{prop:branchA}), or by lifting the $\pi$-adic digits of $T$ one at a time when $\rank T=1$ (Theorem~\ref{thm:hms-lift}). The candidate is certified (Lemma~\ref{lem:check}) and we recurse on a complement (Lemma~\ref{lem:recursion}, Theorem~\ref{thm:qge3}).
\item \emph{Case $q=2$} (Section~\ref{sec:q2}). The flag is no longer determined, so we recover the subgroup. Writing $H=gPg^{-1}$ with $g\equiv I\pmod\m$, two-copy socle shifts strip the digits of $g$ one layer at a time (Proposition~\ref{prop:q2-layer}). A final linear system over $\F_2$ fixes $g$ modulo directions that normalize $P$ (Proposition~\ref{prop:q2-final}, Theorem~\ref{thm:q2}).
\end{enumerate}
One mechanism underlies all three stages. Each probe is chosen so that the unknown $\pi$-adic digit is multiplied into the socle $\pi^{t-1}R\cong k$, where only its residue modulo $\m$ survives. The oracle restricted to such probes is then an abelian hidden subgroup or hidden shift problem over $k$.

\section{Algebraic setting and computational model}\label{sec:prelim}
\subsection{Chain rings and split flags}
\begin{definition}\label{def:fin-chain-ring}
A finite chain ring is a finite commutative local ring whose ideals are linearly ordered by inclusion. Fix a generator $\pi$ of its maximal ideal $\m$, let $t$ be the least positive integer with $\m^t=0$, and put $k=R/\m=\F_q$. For a field we take $\pi=0$ and $t=1$.
\end{definition}
The ideals are $\pi^jR$ for $0\le j\le t$; each nonzero successive quotient has size $q$, so $|R|=q^t$. Write $v(x)=j$ when $x\in\pi^jR\setminus\pi^{j+1}R$, with $v(0)=t$, and $v(\pi^jR)=j$. Then
\[
 R^\times=R\setminus\m,\qquad
 \Ann(\pi^j)=\pi^{t-j}R,\qquad \soc(R)=\pi^{t-1}R.
\]
For $U\le V=R^n$, write $\bar U=(U+\m V)/\m V\le\bar V=k^n$.

\begin{definition}\label{def:split}
A submodule is \emph{split free} if it is a free direct summand of $V$. A flag of type $(n_1,\ldots,n_m)$, with $n_i\ge1$ and $\sum_i n_i=n$, is a chain $\Fl:V=U_0>U_1>\cdots>U_m=0$ of split free submodules with $\rank U_i=\sum_{j>i}n_j$. Its stabilizer $G_{\{\Fl\}}$ is called \emph{parabolic}. The standard flag $\Fl_0$ consists of $U_i=\langle e_{n_1+n_2+\cdots+n_i+1},\dots,e_n \rangle$ where $e_i$ is the $i$-th standard basis vector; its stabilizer $P$, the \emph{standard parabolic} of that type, is block lower triangular.
\end{definition}

\begin{lemma}\label{lem:conjugate}
The stabilizers of flags of a fixed type are exactly the conjugates of the standard parabolic of that type.
\end{lemma}
\begin{proof}
Each quotient $U_i/U_{i+1}$ is a direct summand of the free module $V/U_{i+1}$, hence is free over the local ring $R$. Lifting bases of these quotients gives an adapted basis, in which the stabilizer is $P$. Conversely, $gPg^{-1}$ stabilizes $g\Fl_0$, whose members remain split free of the same ranks.
\end{proof}

\begin{lemma}[Nakayama's lemma]\label{lem:nak}
If $M$ is a finitely generated $R$-module and $M=N+\m M$, then $M=N$.
\end{lemma}

\begin{lemma}[Splitting criterion]\label{lem:splitting}
For $U\le V$ the following are equivalent:
\begin{enumerate}
    \item[(a)] $U$ is split free of rank $r$;
    \item[(b)] $U$ is generated by $r$ elements with linearly independent reductions in $\bar V$;
    \item[(c)] $U$ is pure ($U\cap\pi^jV=\pi^jU$ for all
$j$) and $\dim_k\bar U=r$.
\end{enumerate} In that case $\rank U=\dim_k\bar U$, every complement is split free, and every split free $U'\le U$ is a direct summand of $U$. In particular a nonzero split free submodule is not contained in $\m V$.
\end{lemma}

A proof, including the purity criterion, is given in Appendix~\ref{app:algebra}.

\begin{lemma}\label{lem:reduced-type}
    $\bar{\Fl}$ is a flag in $\bar{V}$ of the same length and type as $\Fl$, with $\dim_k\bar{U}_i=\rank{U_i}$. In particular the type of $\Fl$, and hence $d=\rank U_{m-1}$, is determined by $\bar{\Fl}$.
\end{lemma}
\begin{proof}
For split free $U$, $U/\m U\cong\bar U$ and $\dim_k\bar U=\rank U$. The positive rank gaps in the flag therefore remain positive after reduction.
\end{proof}

\begin{lemma}\label{lem:divisibility}
If $U\le V$ is split free and $0\le j\le t$ then for $x\in V$, $\pi^jx\in U\iff x\in U+\pi^{t-j}V$.
\end{lemma}
\begin{proof}
Fix $V=U\oplus C$ and write $x=u+c$; then $\pi^jx\in U$ iff $\pi^jc=0$ iff $c\in\pi^{t-j}C$.
\end{proof}
In words, multiplying by $\pi^j$ hides any error in $\pi^{t-j}V$. So the condition $\pi^jx\in U$ tests membership $x\in U$ only to precision $t-j$. This is how the valuations $v(\lambda(U_i))$ enter the probe formulas of Section~\ref{sec:graded}.

\begin{lemma}[Densities]\label{lem:density}
Write $c_q(m)=\prod_{i=1}^m(1-q^{-i})$ and $c_\infty=\prod_{i\ge1}(1-2^{-i})>0.288$, so $c_q(m)\ge c_\infty$.
\begin{enumerate}[label=(\roman*)]
\item A uniform $X\in\Mat_m(R)$ is invertible with probability $c_q(m)$.
\item A uniform $X\in\Mat_{m\times l}(R)$, $l\ge m$, is surjective with probability $\prod_{i=l-m+1}^l(1-q^{-i})\ge c_\infty$.
\item $l\ge m$ uniform elements of a free module of rank $m$ generate it with probability $\ge c_\infty$.
\end{enumerate}
In particular none of these probabilities depends on the length $t$.
\end{lemma}
\begin{proof}
Invertibility is detected by the determinant modulo $\m$; generation and surjectivity are detected modulo $\m$ by Lemma~\ref{lem:nak}. Reduction has equal fibres, and the assertions follow by counting full-rank matrices over $k$.
\end{proof}

\subsection{Computational model}\label{sec:model}
The ring is supplied by additive coordinates and multiplication structure constants over $\Z/p^e\Z$, where $p^e=\operatorname{char}R$, together with $\pi,t$ and $q=p^\ell$. We assume a presentation of size polynomial in $\log|R|$ supporting ring arithmetic, valuation, additive cyclic decomposition,
residue-field arithmetic, and a section $\sigma:k\to R$ with $\sigma(0)=0$, all in time $\poly(\log|R|)$. We apply $\sigma$ coordinatewise to vectors and matrices over $k$: for $a\in k$ we write $\widehat a=\sigma(a)$, and call it the \emph{standard lift} of $a$. We also use polynomially many generators of $R^\times$. Appendix~\ref{app:algebra} describes the unit generators and the module algorithms used below. The parameters $e,t,\ell$ are independent: $|R|=p^{\ell t}$ and $1\le e\le t$; $(R,+)$ need not be homocyclic.

\begin{lemma}[Effective operations]\label{lem:effective}
Membership, equality, sums, intersections, splitness, and bases and complements of split free submodules are computable in classical time polynomial in $n,\log|R|$ and the total encoding length of their generators. From a split free flag or submodule $W$, one can also compute polynomial-size
generating sets of $G_{\{\Fl\}}$ and $\{X\in\GL_n(R):(X-I)V\subseteq W\}$.
\end{lemma}
\begin{proof}
The algorithms and group generators are detailed in Appendix~\ref{app:algebra}.
\end{proof}

The HSP oracle $f:G\to S$ is coherent and satisfies $f(x)=f(y)\quad\Longleftrightarrow\quad x^{-1}y\in H.$ Thus $f(xh)=f(x)$ for $h\in H$. Its restriction to any known subgroup $A$ hides $H\cap A$. Inversion converts it to a right-coset oracle with one query and polynomial matrix arithmetic. Output encodings and one oracle call have polynomial size in the stated input parameters.

\begin{convention}[Error and running time]\label{conv:error}
Quantum polynomial time means bounded running time $\poly(n,\log|R|,\log(1/\epsilon))$ and success probability at least $1-\epsilon$. All loops have explicit caps; malformed outputs or timeouts cause a failure return. For at most $N$ adaptive calls with valid promises before the first error, assigning error $\delta/N$ to each bounds the probability of a first erroneous call by $\delta$. Any additional sampling
failure is analyzed in the corresponding ideal execution with exact subroutine answers. Approximate QFTs and state preparations are given a separate error budget: if at most $B$ are used, their individual operator
errors are at most $O(\epsilon/B)$, so the complete computation changes by at most $O(\epsilon)$ in trace distance. Constant-factor budget divisions are chosen so that the total is at most $\epsilon$.
\end{convention}

\begin{problem}[Parabolic HSP]\label{prob:main}
Given the above presentation of $R$ and a coherent oracle hiding a parabolic subgroup $H\le\GL_n(R)$ of unknown type, output generators of $H$.
\end{problem}

\subsection{Faithfulness}\label{sec:faithful}
The distinction between flag and subgroup recovery is governed by
\begin{equation}\label{eq:units}
 \text{there exists }b\in R^\times\text{ with }b-1\in R^\times \quad\Longleftrightarrow\quad q\ge3.
\end{equation}

Equivalently, $q\ge3$ exactly when $k$ has an element other than $0$ and $1$. For $q=2$ every unit is $\equiv1\pmod\m$, so no unit $b$ can make $b-1$ a unit.

\begin{theorem}[Faithfulness for $q\ge3$]\label{thm:faithful}
Suppose $q\ge3$. If $\Fl$ and $\Fl'$ are flags in $V=R^n$ with $G_{\{\Fl\}}\subseteq G_{\{\Fl'\}}$, then every member of $\Fl'$ is a member of $\Fl$. In particular $G_{\{\Fl\}}=G_{\{\Fl'\}}$ implies $\Fl=\Fl'$, and the parabolic subgroup determines the flag.
\end{theorem}

The proof is in Appendix~\ref{app:algebra}. At $q=2$ the conclusion can fail already in dimension two: over $\Z/4\Z$, the distinct lines $Re_1$ and $R(e_1+2e_2)$ have the same stabilizer (Example~\ref{ex:z4}). Proposition~\ref{prop:unfaithful} describes these fibres over every binary chain ring of length at least two. The algorithm must therefore target the subgroup rather than a unique flag.

\begin{definition}[Congruence filtration]\label{def:congruence}
Put $K_j=I+\pi^j\Mat_n(R)$ for $j\ge1$, with $K_j=\{I\}$ for $j\ge t$. These are the kernels of reduction modulo $\pi^j$, hence normal subgroups of $G$. We use the commutator convention $[x,y]=xyx^{-1}y^{-1}$.
\end{definition}
For $t\ge2$ the deepest nontrivial layer is abelian. Since $\pi^{2t-2}=0$, we have $(I+\pi^{t-1}A)(I+\pi^{t-1}B)=I+\pi^{t-1}(A+B)$, and $I+\pi^{t-1}Y\mapsto\bar Y$ identifies $K_{t-1}$ with $(\Mat_n(k),+)$. All probes in Sections~\ref{sec:graded} and~\ref{sec:q2} live in this socle layer, which is why the HSPs they produce are elementary abelian.

\section{Recovering the reduced flag}\label{sec:graded}
Throughout this section $t\ge2$ and $H=G_{\{\Fl\}}$, with $\Fl:V=U_0>\cdots>U_m=0$, terminal member $T=U_{m-1}$ and terminal rank $d=\rank T$. For the one-block flag, $T=V$. We first compute intersections with rank-one matrix families, then use their socle restrictions to recover $\bar\Fl$.

\subsection{Intersection probes}\label{sec:gadgets}
For a unimodular functional $\lambda\in V^*=\Hom_R(V,R)$, meaning $\lambda(V)=R$, put $P_\lambda=\ker\lambda$ and
\[
 G_\lambda=\{I+a\lambda:1+\lambda(a)\in R^\times\},\qquad
 N_\lambda=\{I+a\lambda:a\in P_\lambda\}\cong(R^{n-1},+).
\]
The product is $(I+a\lambda)(I+b\lambda)=I+(a+b+\lambda(b)a)\lambda$. Consequently $G_\lambda$ is a group, $N_\lambda$ is its abelian normal subgroup, and $G_\lambda/N_\lambda\cong R^\times$. Thus $N_\lambda$ consists of the transvections with fixed ``row'' $\lambda$. Its parameter $a$ is a vector, and $I+a\lambda$ moves each $u\in V$ by $\lambda(u)a$. Restricting the oracle to $N_\lambda$ gives an abelian HSP.

\begin{lemma}[Filtered stabilizer formula]\label{lem:gadget}
Let $\lambda\in V^*$ be unimodular and set $v_i=v(\lambda(U_i))$ for $0\le i\le m$, so that $0=v_0\le v_1\le\dots\le v_{m-1}\le v_m=t$. Put
\[
  D_\lambda\ :=\ \bigcap_{i=0}^{m}\bigl(U_i+\pi^{\,t-v_i}V\bigr)  \ =\ \bigcap_{i=1}^{m-1}\bigl(U_i+\pi^{\,t-v_i}V\bigr),
\]
the terms $i=0$ and $i=m$ both being equal to $V$. Then
\begin{enumerate}[label=(\roman*)]
\item $H\cap N_\lambda=\{I+a\lambda:\ a\in P_\lambda\cap D_\lambda\}$ and $H\cap G_\lambda=\{I+a\lambda:\ a\in D_\lambda,\ 1+\lambda(a)\in R^\times\}$;
\item $D_\lambda$, and hence each of the two intersections, is a function of $\lambda$, of the valuations $v_0,\dots,v_m$, and of the residues $U_i\bmod\pi^{\,t-v_i}V$ alone. 
\end{enumerate}
\end{lemma}
\noindent The integer $v_i$ measures how deep $\lambda(U_i)$ lies in the chain of ideals of $R$. A member with $v_i=0$ ($\lambda$ surjective on $U_i$) imposes the exact condition $a\in U_i$. A member with $0<v_i<t$ imposes it only modulo $\pi^{t-v_i}V$, and a member with $v_i=t$ imposes nothing. So $H\cap N_\lambda$ sees each $U_i$ only to precision $t-v_i$, and part~(ii) says that it sees nothing more.

\begin{proof}
(i) For $X=I+a\lambda$ and $u\in U_i$ we have $Xu=u+\lambda(u)a$, so $XU_i\subseteq U_i$ iff $\lambda(U_i)a\subseteq U_i$, i.e.\ $\pi^{v_i}a\in U_i$, i.e.\ $a\in U_i+\pi^{\,t-v_i}V$ by Lemma~\ref{lem:divisibility}; and $XU_i\subseteq U_i$ forces $XU_i=U_i$ because $X$ is invertible and $U_i$ is finite. Intersect over $i$ and impose $a\in P_\lambda$, respectively invertibility of $I+a\lambda$, which by $\det(I+a\lambda)=1+\lambda(a)$ is the stated unit condition. For the two trivial terms, $v_0=0$ gives $U_0+\pi^tV=V$ and $v_m=t$ gives $0+\pi^0V=V$. (ii) is immediate, since $U_i+\pi^{\,t-v_i}V$ depends on $U_i$ only through its class modulo $\pi^{\,t-v_i}V$.
\end{proof}

Call $\lambda$ \emph{transverse} if $\lambda(T)=R$. Equivalently, $\bar\lambda$ does not vanish on $\bar T$. Then every $v_i$ with $i<m$ is $0$, so the smallest member $T$ determines $D_\lambda$ by itself.

\begin{corollary}\label{cor:transverse}
If $\lambda$ is transverse then $D_\lambda=T$, so
\[
  H\cap N_\lambda=\{I+a\lambda:a\in T\cap P_\lambda\}\cong T\cap P_\lambda ,
\]
which is split free of rank $d-1$, and
\[
  H\cap G_\lambda=\{I+a\lambda:\ a\in T,\ 1+\lambda(a)\in R^\times\} ,
\]
a subgroup of $G_\lambda$ whose coefficient set has $R$-span $T$ provided $q\ge3$.
\end{corollary}

\begin{proof}
Since $\lambda(T)=R$ and $T\subseteq U_i$ for $i<m$, we have $\lambda(U_i)=R$, so $v_i=0$ for all $i<m$ and $D_\lambda=T$. Because $\lambda(T)=R$, we may fix $\tau\in T$ with $\lambda(\tau)=1$. Then $r\mapsto r\tau$ splits $\lambda|_T\colon T\to R$, and
\[
  T=(T\cap P_\lambda)\oplus R\tau,\qquad a=\bigl(a-\lambda(a)\tau\bigr)+\lambda(a)\tau .
\]
The first assertion follows from the additive product on $N_\lambda$. For the second, note that the coefficient set $\{a\in T:1+\lambda(a)\in R^\times\}$ contains $T\cap P_\lambda$. If $q\ge3$, choose $c\in R^\times$ with $1+c\in R^\times$. Then $c\tau$ also lies in the coefficient set, since $1+\lambda(c\tau)=1+c$. As $c$ is a unit, the $R$-span of the coefficient set is $(T\cap P_\lambda)+R\tau=T$.
\end{proof}

\begin{lemma}[Transversality is generic]\label{lem:generic}
For $\lambda$ uniform among unimodular elements of $V^*$,
\[
  \Pr[\lambda\text{ not transverse}]   =\frac{q^{-d}-q^{-n}}{1-q^{-n}}   =q^{-d}\cdot\frac{1-q^{\,d-n}}{1-q^{-n}}   \ <\ q^{-d}\ \le\ \frac1q .
\]
\end{lemma}

\begin{proof}
In coordinates adapted to $T$, nontransversality says that the first $d$ coefficients of $\lambda$ lie in $\m$. Before conditioning on unimodularity this has probability $q^{-d}$; the excluded event that all $n$ coefficients lie in $\m$ has probability $q^{-n}$.
\end{proof}

\begin{proposition}[Terminal rank at least two]\label{prop:branchA}
Suppose $\rank T\ge2$ and draw $\lambda$ uniformly among unimodular elements of $V^*$. An abelian HSP in $N_\lambda\cong(P_\lambda,+)\cong R^{\,n-1}$ returns $H\cap N_\lambda$ in time $\poly(n,\log|R|)$. With probability
\[
  1-\frac{q^{-d}-q^{-n}}{1-q^{-n}}\ >\ 1-q^{-d}\ \ge\ 1-q^{-2}\ \ge\ \frac34
\]
the draw is transverse, and in that case
\[
  \sum_{X\in H\cap N_\lambda}(X-I)V\ =\ T\cap P_\lambda
\]
is a nonzero split free submodule of $T$, of rank $d-1$.
\end{proposition}

\begin{proof}
$P_\lambda$ is split free of rank $n-1$ because $V/P_\lambda\cong R$ is free, so $(N_\lambda,+)\cong(R^{n-1},+)$ is a finite abelian group of known structure and the abelian HSP applies at the stated cost; only the
fact that $f$ restricted to a subgroup hides the intersection is used, so normality of $N_\lambda$ is not needed. The probability is Lemma~\ref{lem:generic}. For a transverse draw, Corollary~\ref{cor:transverse} gives $H\cap N_\lambda=\{I+a\lambda:a\in T\cap P_\lambda\}$, and $(I+a\lambda-I)V=a\lambda(V)=aR$ since $\lambda$ is unimodular, so the sum is $T\cap P_\lambda$, of rank $d-1\ge1$.
\end{proof}

The Frobenius duality and constant-density estimates behind the DIKQS \cite{DIKQS14} Fourier procedures extend to $R$, but one transported rank-one procedure has repetition factor $(1-1/q)^{-(n-1)}$ (Appendix~\ref{sec:transport}). The new lifting procedures avoid this.

\subsection{Socle probes}\label{sec:reduced}

\begin{definition}[Socle layers]\label{def:socle}
Let $t\ge2$ and let $\mu\in V^*$ be unimodular. Put
\[
  S_\mu=\{\,I+\pi^{\,t-1}a\mu\ :\ a\in V\,\}\ \le\ G .
\]
\end{definition}

\begin{lemma}[Socle layers]\label{lem:socle}
Let $t\ge2$ and let $\mu\in V^*$ be unimodular, and write $q=p^\ell$. Then:
\begin{enumerate}[label=(\roman*)]
\item $S_\mu=G_\mu\cap K_{t-1}$, and $a\mapsto I+\pi^{\,t-1}a\mu$ induces a group isomorphism $(\bar V,+)\xrightarrow{\ \sim\ }S_\mu$; in particular $S_\mu\cong(\Z/p\Z)^{\ell n}$ is elementary abelian of order $q^n$, and it is generated by the $\ell n$ explicitly computable elements
\[
 I+\pi^{\,t-1}\sigma(b_r)e_a\mu,  \qquad 1\le a\le n,\ 1\le r\le \ell,
\]
where $b_1,\dots,b_\ell$ is an $\F_p$-basis of $k$ and $e_1,\dots,e_n$ is the standard basis of $V$;
\item let $i_0(\mu)$ be the largest index $i$ with $\mu(U_i)=R$. It is well defined, and $\mu(U_i)=R$ iff $\bar U_i\not\subseteq\ker\bar\mu$, so $i_0(\mu)$ depends only on $\bar\mu$. Moreover
\[
  H\cap S_\mu=\{\,I+\pi^{\,t-1}a\mu:\bar a\in\bar U_{i_0(\mu)}\,\} \ \cong\ \bar U_{i_0(\mu)} .
\]
\end{enumerate}
\end{lemma}
\noindent The socle probe is the depth-$(t-1)$ version of $G_\mu$. Because its coefficients are multiplied by $\pi^{t-1}$, only $\bar a$ and $\bar\mu$ matter. The intersection with $H$ is therefore governed by the reduced flag alone. It returns the smallest reduced member $\bar U_{i_0(\mu)}$ on which $\bar\mu$ is not identically zero.

\begin{proof}
Since $\pi^{2t-2}=0$,
\[
 (I+\pi^{t-1}a\mu)(I+\pi^{t-1}b\mu) =I+\pi^{t-1}(a+b)\mu.
\]
A unit coordinate of $\mu$ shows that the kernel of the parameter map $V\to S_\mu$ is $\m V$, and that $G_\mu\cap K_{t-1}=S_\mu$. This proves (i), including the generators obtained from the $\F_p$-basis $\{b_re_a\}$ of $k^n$.

For (ii), $i_0(\mu)$ exists because $\mu(V)=R$, and $\mu(U_i)=R$ exactly when $\bar\mu$ is nonzero on $\bar U_i$. An element $I+\pi^{t-1}a\mu$ preserves $U_i$ exactly when $\pi^{t-1}\mu(U_i)a\subseteq U_i$. If $\mu(U_i)\subseteq\m$ this is automatic; otherwise Lemma~\ref{lem:divisibility} makes it equivalent to $\bar a\in\bar U_i$. Intersecting these nested conditions yields $\bar a\in\bar U_{i_0(\mu)}$.
\end{proof}

\begin{proposition}\label{prop:reduced}
Let $R$ be a finite chain ring of length $t\ge2$ and let $0<\epsilon<1$. There is a quantum algorithm running in time $\poly(n,\log|R|,\log(1/\epsilon))$ which, for every $q$ and every $n$ and with no further hypotheses, outputs the reduced flag
\[
  \bar V=\bar U_0>\bar U_1>\dots>\bar U_m=0
\]
in $\bar V=k^n$, together with its type $(n_1,\dots,n_m)$, with failure probability at most $\epsilon$.
\end{proposition}
\noindent\emph{Idea.} The flag is built from the bottom. Suppose $W=\bar U_{j+1}$ has been found, and take $\mu$ with $\bar\mu$ vanishing on $W$. By Lemma~\ref{lem:socle}(ii) the answer is a member containing $\bar U_j$, and with probability at least $1/2$ it is $\bar U_j$ itself. Intersecting $r$ such answers gives $\bar U_j$ unless every draw also vanishes on $\bar U_j$.

\begin{proof}
Let $r=\lceil\log_2(2n/\epsilon)\rceil$. Maintain a recovered subspace $W$, initially $0$. While $W\ne\bar V$, sample $r$ independent nonzero functionals uniformly from the annihilator $W^\circ:=\{\lambda\in \bar V^*: \lambda|_W=0\}$ of $W$ in $\bar V^*$, lift each to a unimodular $\mu$, and solve the abelian HSP in $S_\mu$ with error at most $\epsilon/(2nr)$. Convert each answer to the $k$-span of its coefficient vectors and let
$W_{\rm new}$ be the intersection of those spans. Require $W<W_{\rm new}\le\bar V$, append $W_{\rm new}$, and continue. Stop after at most $n$ rounds, reporting failure if strict progress or termination fails. Reverse the resulting list to obtain the flag and read its type from dimension gaps.

Analyze first the ideal execution with exact intersection answers. If the current member is $W=\bar U_{j+1}$, every sampled functional kills $W$, and Lemma~\ref{lem:socle} returns a flag member containing $\bar U_j$. It equals $\bar U_j$ unless it also kills $\bar U_j$. For a uniform nonzero element of $W^\circ$, the latter probability is at most $q^{-(\dim\bar U_j-\dim W)}\le1/2$. Thus the next member is missed with probability at most $2^{-r}$. Every ideal round makes strict progress, even if it skips a member, so there are at most $n$ rounds and the probability of any omission is at most $n2^{-r}\le\epsilon/2$.

Couple the actual and ideal executions up to the first incorrect HSP answer. Before that error they have the same history, so each call has its stated promise and error bound. At most $nr$ calls give discrepancy probability at most $\epsilon/2$; adding the ideal sampling error proves
the result. This argument does not condition probe distributions on future success events. The caps bound all execution paths. Each query forms a matrix over $R$, giving the claimed $\poly(n,\log|R|,\log(1/\epsilon))$ gate cost.
\end{proof}

A successful call also supplies a basis change making the reduced flag standard. Lemma~\ref{lem:q2-standard} then gives $H=gPg^{-1}$ with $g\in K_1=I+\pi\Mat_n(R)$ and $P$ the known standard parabolic. The following sections recover the remaining digits.

\section{Rank-one lifting}\label{sec:hmslift}
Assume $t\ge2$, $q\ge3$, and that the reduced terminal line is known. Write $l=n-1$ in this section. After a known basis change the terminal member has the unique normalized form
\begin{equation}\label{eq:normalized-line}
 T=R(e_1+y),\qquad y\in\m R^l.
\end{equation}
Since $\sigma(k)$ is a set of representatives containing $0$, $y$ has a unique expansion
$y=\sum_{i=1}^{t-1}\pi^i\sigma(s_i)$ with $s_i\in k^l$. For $1\le j\le t$ write
\begin{equation}\label{part-y}
    y_{<j}=\sum_{i=1}^{j-1}\pi^i\sigma(s_i),
\end{equation}
so that $y\equiv y_{<j}\pmod{\pi^jR^l}$. Once the digits $s_1,\dots,s_{j-1}$ are known,
conjugating by $C_j=\begin{psmallmatrix}1&0\\-y_{<j}&I_l\end{psmallmatrix}$ maps $T$ to
\begin{equation}\label{eq:Tj}
 T_j=R(e_1+\pi^jz_j),\qquad z_j\in R^l,\quad s_j=\bar z_j\in k^l,\qquad 1\le j<t.
\end{equation}
Write $H_j,f_j$ for the conjugated subgroup and oracle.

\paragraph{Standing notation for this section.}
Here $l=n-1$, and matrices are written in blocks of sizes $1$ and $l$. The terminal line is $T=R(e_1+y)$ with $\pi$-adic digits $s_1,\dots,s_{t-1}\in k^l$. At stage $j$ the digits $s_1,\dots,s_{j-1}$ are known and have been conjugated away. The current line is $T_j=R(e_1+\pi^jz_j)$, its unknown leading digit is $s_j=\bar z_j$, and $H_j,f_j$ are the subgroup and oracle after this conjugation. The section $\sigma$ and the notation $\widehat a=\sigma(a)$ are from Section~\ref{sec:model}.

\subsection{One lifting layer}
Use the fixed section $\sigma$ coordinatewise, writing $\widehat a=\sigma(a)$. For $x\in k^l$ and $a\in k$, define the query matrix
\begin{equation}\label{eq:Gja}
 G_j(x,a)=
 \begin{pmatrix}1+\pi^{t-1-j}\widehat a&0\\
 \pi^{t-1}\widehat x&I_l\end{pmatrix}.
\end{equation}
Think of $x\in k^l$ as a test vector and $a\in k$ as a scalar. The diagonal entry has depth $t-1-j$, chosen so that its product with the unknown $\pi^jz_j$ (depth $j$) lands exactly in the socle $\pi^{t-1}R$. There it is compared with the lower-left block $\pi^{t-1}\widehat x$. The section need not preserve addition or multiplication. These matrices need not form a group; the HSP below is on their
parameter group. For the coset calculations use the containing group
\[
 Q=\left\{M(\beta,c)=
 \begin{pmatrix}\beta&0\\\pi^{t-1}\widehat c&I_l\end{pmatrix} :\beta\in R^\times,\ c\in k^l\right\},
\]
whose operations are
\begin{equation}\label{eq:grouplaw}
\begin{aligned}
 M(\beta,c)M(\beta',c')&=M(\beta\beta',c\bar\beta'+c'),\\
 M(\beta,c)^{-1}&=M(\beta^{-1},-c\bar\beta^{-1}).
\end{aligned}
\end{equation}
For $1\le r<t$ the well-defined map
\begin{equation}\label{eq:digitmap}
 b_r:1+\pi^rR\to(k,+),\qquad b_r(1+\pi^ru)=\bar u
\end{equation}
is a surjective homomorphism: the cross term in a product contributes $\pi^ruu'\in\m$. Separately put
\begin{equation}\label{eq:digitmap0}
 b_0:R^\times\to k,\qquad b_0(\beta)=\bar\beta-1.
\end{equation}
Unlike $b_r$ for $r\ge1$, this last map is not an additive homomorphism.

\begin{lemma}[Membership]\label{lem:layer-identity}
Let $1\le j\le t-1$ and let $s_j$ be as in \eqref{eq:Tj}.
\begin{enumerate}[label=(\roman*)]
\item If $j<t-1$, then $G_j(x,a)\in\GL_n(R)$ for every $a\in k$. If $j=t-1$, it is invertible exactly when $a\ne-1$.
\item More generally, put $r=t-1-j$ and let $M(\beta,c)\in Q$, where $\beta\in1+\pi^rR$ if $r\ge1$ and $\beta\in R^\times$ if $r=0$, and $c\in k^l$. Then
\begin{equation}
  M(\beta,c)\in H_j
  \iff
  c=b_r(\beta)\,s_j ,
\label{eq:layer-membership}
\end{equation}
with $b_r$ as in \eqref{eq:digitmap} for $r\ge1$ and as in \eqref{eq:digitmap0} for $r=0$. In particular, for admissible $a$, $G_j(x,a)\in H_j$ if and only if $x=a\,s_j$.
\end{enumerate}
\end{lemma}
\noindent Formula~\eqref{eq:layer-membership} says that $M(\beta,c)$ fixes $T_j$ exactly when its lower-left block $\pi^{t-1}\widehat c$ matches the socle term $(\beta-1)\pi^jz_j=\pi^{t-1}uz_j$. Here $\beta=1+\pi^ru$ with $\bar u=b_r(\beta)$, and $s_j=\bar z_j$. The membership condition is therefore the bilinear relation $c=(\text{digit of }\beta)\cdot s_j$ over $k$.

\begin{proof}
The determinant of $G_j(x,a)$ is $1+\pi^{t-1-j}\widehat a$, proving (i). For (ii), write $\beta=1+\pi^ru$, with $u=\beta-1$ if $r=0$,
and put $v_j=e_1+\pi^jz_j$. The first coordinate of $M(\beta,c)v_j$ is $\beta$, so $M(\beta,c)$ stabilizes $T_j$ exactly when
$\pi^{t-1}\widehat c=(\beta-1)\pi^jz_j=\pi^{t-1}uz_j$. Since $\Ann(\pi^{t-1})=\m$, this is $c=\bar u s_j=b_r(\beta)s_j$. It is also sufficient for membership in the full flag stabilizer, because then
\begin{equation}\label{eq:layer-column}
 M(\beta,c)-I=(\pi^ruv_j)e_1^*,
\end{equation}
so $(M-I)V\subseteq T_j$, which is contained in every nonzero flag member.
\end{proof}

\begin{lemma}[Intermediate layers]\label{lem:layer-intermediate}
Let $1\le j<t-1$ and put $F_j(x)=f_j(G_j(x,0))$. Then $\{G_j(x,0):x\in k^l\}$ is a subgroup isomorphic to $k^l$, the function
$F_j$ is injective, and for every $a\in k$ and $x\in k^l$
\begin{equation}\label{eq:layer-oracle}
  f_j\bigl(G_j(x,a)\bigr)=F_j\bigl(x-a\,s_j\bigr).
\end{equation}
The shift parameter $a$ ranges over all of $k$.
\end{lemma}
\noindent In other words, $(x,a)\mapsto f_j(G_j(x,a))$ is constant exactly on cosets of the line $\{(as_j,a):a\in k\}$ in $(k^l\times k,+)$, which is an abelian HSP. The point is that for $r=t-1-j\ge1$ the digit map $b_r$ is additive, so composing two queries adds their scalars $a$.

\begin{proof}
Since $\sigma(0)=0$ we have $G_j(x,0)=M(1,x)$, so \eqref{eq:grouplaw} gives $G_j(x,0)G_j(x',0)=G_j(x+x',0)$ and $G_j(x,0)^{-1}=G_j(-x,0)$: this family is a subgroup isomorphic to $k^l$. Write $r=t-1-j$, $\alpha=1+\pi^r\widehat a$ and $\alpha'=1+\pi^r\widehat{a'}$, and note $\bar\alpha=\bar{\alpha'}=1$ because $r\ge1$. Then \eqref{eq:grouplaw} gives
\[
  G_j(x,a)^{-1}G_j(x',a')=M\bigl(\alpha^{-1}\alpha',\,x'-x\bigr).
\]
Here $\alpha^{-1}\alpha'$ lies in the group $1+\pi^rR$, and since $r\ge1$ the map $b_r$ of \eqref{eq:digitmap} is a homomorphism, so
$b_r(\alpha^{-1}\alpha')=-b_r(\alpha)+b_r(\alpha')=a'-a$. By Lemma~\ref{lem:layer-identity}(ii) this matrix lies in $H_j$ if and only
if $x'-x=(a'-a)s_j$. Taking $a'=0$ and $x'=x-a s_j$ shows that $G_j(x,a)$ and $G_j(x-as_j,0)$ lie in the same left coset of $H_j$,
which is \eqref{eq:layer-oracle}; taking $a=a'=0$ shows $G_j(x,0)^{-1}G_j(x',0)\in H_j$ iff $x=x'$, which is injectivity of $F_j$.
\end{proof}

\begin{lemma}[Final layer]\label{lem:layer-final}
Let $j=t-1$, write $h=\bar\beta$ for $\beta\in R^\times$, so that the admissible query coefficients are $h=1+a$ with $a\ne-1$, and abbreviate $s=s_{t-1}$, $\Gamma=Q$. Then:
\begin{enumerate}[label=(\roman*)]
\item $\Gamma$ is a subgroup of $\GL_n(R)$ and $M(\beta,c)\mapsto(c,\bar\beta)$ is a surjective homomorphism $\Gamma\to k^l\rtimes k^\times$, the target having the product $(x,h)(x',h')=(xh'+x',\,hh')$. Moreover $\Gamma\cap H_{t-1}=\{M(\beta,c):c=(\bar\beta-1)s\}$ is the full preimage of $\{(x,h):x=(h-1)s\}$, so all coset computations may be carried out in $k^l\rtimes k^\times$;
\item two elements of $\Gamma$ lie in the same left coset of $\Gamma\cap H_{t-1}$ if and only if they have the same value of
$h^{-1}(x+s)\in k^l$;
\item consequently, setting $\Phi(w)=f_{t-1}(G_{t-1}(w-s,0))$, the function $\Phi$ is injective and for all $w\in k^l$, $h\in k^\times$
\begin{equation}\label{eq:layer-final-oracle}
  f_{t-1}\bigl(G_{t-1}(hw,\,h-1)\bigr)=\Phi\bigl(w+h^{-1}s\bigr).
\end{equation}
Equivalently, writing $c=-h^{-1}$, the family $\{\,w\mapsto f_{t-1}(G_{t-1}(hw,h-1))\,\}$ is of the form $\Phi(w-c\,s)$ with $c$ ranging over $k^\times$.
\end{enumerate}
\end{lemma}
\noindent At the last digit ($r=0$) the diagonal entry ranges over all of $R^\times$, and $b_0$ is no longer additive. The relevant group is the affine group $k^l\rtimes k^\times$, acting on $k^l$ by $w\cdot(x,h)=wh+x$. Part~(i) says that $\Gamma\cap H_{t-1}$ is the stabilizer of the point $-s$, since $-sh+x=-s$ iff $x=(h-1)s$. Part~(ii) labels its cosets by the image of that point. The hidden subgroup is thus a nonabelian point stabilizer, and part~(iii) recasts it as $q-1$ shifted copies of a single injective function.

\begin{proof}
(i) The product law is \eqref{eq:grouplaw}, and the displayed map is a homomorphism onto $k^l\rtimes k^\times$ because $\beta\mapsto\bar\beta$ is one. The intersection is Lemma~\ref{lem:layer-identity}(ii) with $r=0$, where $b_0(\beta)=\bar\beta-1$. Its description shows that the kernel $\{M(\beta,0):\beta\in1+\m\}$ of the map is contained in $H_{t-1}$, so $\Gamma\cap H_{t-1}$ is the full preimage of $\{(x,h):x=(h-1)s\}$ and descends to the quotient.

(ii) From (i), $(x,h)^{-1}=(-xh^{-1},h^{-1})$, so
\[
  (x,h)^{-1}(x',h')=\bigl(x'-xh^{-1}h',\ h^{-1}h'\bigr),
\]
which lies in $\Gamma\cap H_{t-1}$ iff $x'-xu=(u-1)s$ with $u=h^{-1}h'$, i.e.\ iff $x'+s=u(x+s)$, i.e.\ iff $h'^{-1}(x'+s)=h^{-1}(x+s)$.

(iii) By (ii) the value $f_{t-1}(G_{t-1}(x,h-1))$ is an injective function of $h^{-1}(x+s)$; at $h=1$, $x=w-s$ this function is $\Phi$, so $\Phi$ is injective. Substituting $x=hw$ gives $h^{-1}(hw+s)=w+h^{-1}s$, which is \eqref{eq:layer-final-oracle}. As
$h$ ranges over $k^\times$ so does $h^{-1}$, hence so does $c=-h^{-1}$.
\end{proof}

\begin{remark}\label{rem:reparam}
For $F_0(x)=f_{t-1}(G_{t-1}(x,0))$, the final-layer identity is
$f_{t-1}(G_{t-1}(x,a))=F_0((1+a)^{-1}(x-as))$.
The substitution $x=hw$ and $\Phi(w)=F_0(w-s)$ produce genuine shifts
of one common injective function, as in \eqref{eq:layer-final-oracle}.
\end{remark}

\subsection{Recovering an intermediate digit}

\begin{proposition}[Intermediate lifting]\label{prop:intermediate-lift}
Let $1\le j<t-1$. Given the first $j-1$ digits of the rank-one terminal
member, the next digit $s_j\in k^l$ can be recovered in quantum time
$\poly(n,\log|R|)$.
\end{proposition}

\begin{proof}
Lemma~\ref{lem:layer-intermediate} makes
$(x,a)\mapsto f_j(G_j(x,a))=F_j(x-as_j)$ an oracle for
\[
 L_{s_j}=\{(as_j,a):a\in k\}\le(k^l\times k,+).
\]
An abelian HSP returns additive generators of this subgroup.
Any nonzero returned element has the form $(as_j,a)$ with $a\ne0$;
divide its first component by $a$ to recover $s_j$.
The group computations are over $k$, and the query matrices are
computed over $R$, giving the stated gate cost.
\end{proof}

\subsection{The final digit}
Lemma~\ref{lem:layer-final} reduces the remaining task to $q-1$ shifted
copies indexed by $k^\times$. We work directly over the additive space
$k^l$, including when $q$ is not prime.

\begin{proposition}[Finite-field $(q-1)$-shift]\label{prop:finite-field-shift}
Let $k=\F_q$ with $q\ge3$, let $F:k^l\to S$ be injective, and suppose we
have quantum oracle access to
\begin{equation}
 \Psi_s(w,c)=F(w-c\,s),
 \qquad w\in k^l,\quad c\in k^\times,
 \label{eq:ffshift}
\end{equation}
for an unknown $s\in k^l$. Then $s$ can be recovered in quantum time
$\poly(l,\log q)$.
\end{proposition}

\begin{proof}
Choose distinct $c_0,c_1\in k^\times$. Injectivity of $F$ gives the
exact verifier
\begin{equation}\label{eq:shift-verify}
 z=s\quad\Longleftrightarrow\quad
 \Psi_s(0,c_0)=\Psi_s((c_1-c_0)z,c_1).
\end{equation}
First test $z=0$, and henceforth assume $s\ne0$.

Fix a nontrivial additive character $\chi$ of $k$. Querying on a uniform
superposition of $(w,c)\in k^l\times k^\times$ and measuring the output
leaves $(q-1)^{-1/2}\sum_{c\ne0}|y+cs\rangle|c\rangle$.
(The measured value $F(y)$ fixes $y=w-cs$ by injectivity of $F$.)
Fourier transform the first register and measure it. The outcome
$u\in k^l$ is uniform, and the remaining state, up to a global phase, is
\begin{equation}\label{eq:mu-state}
 \mu_{\ell(u)}=\frac1{\sqrt{q-1}}\sum_{c\ne0}
       \chi(c\ell(u))|c\rangle,\qquad \ell(u)=u\cdot s.
\end{equation}
The unknown $s$ now enters only through the scalar $\ell(u)$. It therefore suffices to learn the nonzero linear form $\ell:k^l\to k$, and the two regimes below differ only in how $\ell(u)$ is extracted from $\mu_{\ell(u)}$. Note that $\mu_\theta$ is the Fourier state $\kappa_\theta$ defined below with its $c=0$ term removed.

\emph{Small $q$: $q\le(l+1)^2$.}
Put $\kappa_\theta=q^{-1/2}\sum_{c\in k}\chi(c\theta)|c\rangle$.
Project the remaining register onto the known unit vector
\begin{equation}\label{eq:eta}
 \eta=\frac{\kappa_0-\kappa_1}{\sqrt2}
      =\frac1{\sqrt{2q}}\sum_{c\in k}(1-\chi(c))|c\rangle.
\end{equation}
It is supported on $k^\times$ and can be prepared by a QFT applied to
$(|0\rangle-|1\rangle)/\sqrt2$; reversing that preparation implements
the projection. Character orthogonality gives
\[
 |\langle\eta,\mu_\theta\rangle|^2
 =\begin{cases}q/(2(q-1)),&\theta\in\{0,1\},\\0,&\text{otherwise}.
 \end{cases}
\]
Since $\ell\ne0$, acceptance has probability $1/(q-1)$, and the
accepted $u$ is uniform on $\ell^{-1}(\{0,1\})$.
So the projection postselects on $\ell(u)\in\{0,1\}$. Learning $\ell$ from uniform samples of the preimage of a fixed subset of $k$, here of size $d=2$, is the problem of learning a linear function from subset \cite{IPS18}.
This is an $\operatorname{LFS}(q,l,2)$ instance.
The algorithm of \cite[Thms.~1.3 and~3.2]{IPS18} uses polynomially many
samples and polynomial expected time, and recovers $\ell$ up to
nonzero scalar with probability at least $1/2$.
Its LFS theorem applies to prime-power $q$ with $2\le d\le q-1$,
so $d=2$ is valid for every $q\ge3$.
Enumerate the at most $q-1$ scalar multiples of its output and apply
\eqref{eq:shift-verify}.

For bounded running time, let $M=\poly(l,\log q)$ be the prescribed
sample count and use $B=C(q-1)(M+1)$ raw trials.
For a suitable absolute $C$, a binomial tail bound makes the probability
of fewer than $M$ acceptances at most $1/16$.
Use the first $M$ accepted samples, whose joint conditional distribution
is the required independent LFS distribution; the acceptance indicators
are independent of their accepted values.
Also cap the randomized LFS postprocessing at eight times a polynomial
upper bound for its expected running time. Markov's inequality bounds
timeout probability by $1/8$.
One capped attempt therefore succeeds with probability at least
$1/2-1/16-1/8>1/4$. Its cost is polynomial because $q\le(l+1)^2$.

\emph{Large $q$: $q>(l+1)^2$.}
An inverse QFT on \eqref{eq:mu-state} returns $\ell(u)$ with probability
$(q-1)/q$, independently of $u$, because $\mu_\theta$ differs from $\kappa_\theta$ only by the missing $c=0$ term. With $m$ independent samples, the
probability that all reported values are correct and the $u$'s are
linearly independent is
\[
 \left(\prod_{i=1}^l(1-q^{-i})\right)(1-1/q)^l
 \ge c_\infty(1-l/q)\ge c_\infty/2.
\]
Solve the resulting linear system and verify its candidate by
\eqref{eq:shift-verify}; singular systems count as failed attempts.

In either regime, $O(\log(1/\epsilon))$ independent capped attempts
reduce the ideal failure probability to $\epsilon/2$, with no false
acceptance. Implementing the QFTs and preparations so that the entire
bounded computation has trace-distance error at most $\epsilon/2$
proves the stated bound. This includes the rejected trials and thus
requires no separate conditioning estimate for approximate postselection.
\end{proof}

\begin{remark}\label{rem:whereStrength}
The chosen threshold is convenient rather than sharp. Direct Fourier
recovery has constant success for $q=\Omega(l)$ and inverse-polynomial
success for $q=\Omega(l/\log l)$, with the corresponding repetition
overhead. The LFS branch is polynomial whenever $q=\poly(l)$.
\end{remark}

\subsection{Completing the lift}

\begin{theorem}[Rank-one lifting over finite chain rings]\label{thm:hms-lift}
Let $R$ be a finite chain ring of length $t\ge2$ with residue field
$k=\F_q$, $q\ge3$. Given the reduced terminal line of a hidden
parabolic subgroup of $\GL_n(R)$ whose terminal member $T$ has rank one,
$T$ can be recovered exactly in quantum time $\poly(n,\log|R|)$.
\end{theorem}

\begin{proof}
After moving the known reduction to $ke_1$, initialize $y_{<1}=0$.
Suppose $y-y_{<j}\in\pi^jR^l$ where $y_{<j}$ is as in (\ref{part-y}). Conjugation by $ C_j=\begin{pmatrix}1&0\\-y_{<j}&I_l\end{pmatrix}$
puts the terminal line in the form \eqref{eq:Tj}.
The digit $s_j$ is well defined: any two choices of $z_j$ differ by
$\pi^{t-j}R^l$, whose reduction is zero.
Use Proposition~\ref{prop:intermediate-lift} for $j<t-1$, and
Lemma~\ref{lem:layer-final} with Proposition~\ref{prop:finite-field-shift}
for $j=t-1$. Set $y_{<j+1}=y_{<j}+\pi^j\sigma(s_j)$.
Then $y-y_{<j+1}\in\pi^{j+1}R^l$, and after $t-1$ stages it is zero.

Give each stage error $\epsilon/(t-1)$.
A first-error union bound is valid because every stage has its stated
promise before the first incorrect digit. Since $t\le\log_2|R|$ and
every stage has bounded polynomial time, the total satisfies
Convention~\ref{conv:error}. Intermediate candidates can also be
checked by $f_j(G_j(z,1))=f_j(I)$; the last digit uses
\eqref{eq:shift-verify}.
\end{proof}

\begin{remark}[Length two and the binary exception]\label{rem:hms-relation}
\label{cor:p2-allp}
For $R=\Z/p^2\Z$ with $p$ odd, the only unknown digit is
$\operatorname{HMS}(p,n-1,p-1)$ with shift set $\F_p^\times$.
For $q=2$, that set has one element, so the final query family alone
contains no shift information. Section~\ref{sec:q2} instead uses shifts
indexed by elements of the whole reduced parabolic.
\end{remark}

\section{Full flag reconstruction for \texorpdfstring{$q\ge3$}{q >= 3}}\label{sec:main}
We combine the reduced-flag and rank-one routines with a verified recursion. The following criterion ensures that every accepted reduction of dimension is legitimate.

\begin{lemma}[Checking: \ {\cite[Lem.~10]{DIKQS14}}]\label{lem:check}
Assume $q\ge3$. Let $0<W<V$ be split free with complement $W'$ and put
$\mathcal L=\{X\in G:(X-I)V\subseteq W\}$ and
$\mathcal L'=\{X\in G:(X-I)V\subseteq W',(X-I)W'=0\}$. Then
$W\subseteq T$ iff $\mathcal L\subseteq H$; and if $W\subseteq T$ then
$W=T$ iff $H\cap\mathcal L'=\{I\}$.
\end{lemma}
\noindent Recall that $T$ is the terminal member of the hidden flag. $\mathcal L$ consists of the elements moving every vector by an element of $W$, and it lies in every stabilizer of a flag whose members all contain $W$. The first criterion is a containment test, checkable on the generators of Lemma~\ref{lem:effective}. The second detects whether $W$ already exhausts $T$.

\begin{proof}
If $W\subseteq T$, every $X\in\mathcal L$ preserves each flag member.
Conversely, choose $w\in W\setminus T$, a unimodular $\tau\in T$, and
$\ell\in V^*$ with $\ell(\tau)=1$.
For a unit $c$, the map $X_c=I+cw\ell$ has image deviation in $W$ and
$X_c\tau=\tau+cw\notin T$. Its determinant is $1+c\ell(w)$.
This is a unit automatically when $\ell(w)\in\m$; otherwise choose
$\bar c\ne-\overline{\ell(w)}^{-1}$, possible because $q\ge3$.

For the second assertion,
$\mathcal L'\cong\Hom_R(W,W')$.
If $W<T$, the decomposition $T=W\oplus(T\cap W')$ supplies
nonzero maps $W\to T\cap W'$ and hence nonidentity elements of
$H\cap\mathcal L'$.
If $W=T$, write $v=w+w'$. An element of $H\cap\mathcal L'$ fixes $w'$
and has $Xw-w\in W\cap W'=0$, so it is the identity.
\end{proof}

The first criterion is checked on generators of $\mathcal L$
(Lemma~\ref{lem:effective}). It is specific to $q\ge3$:
the two lines in Example~\ref{ex:z4} have indistinguishable stabilizers.

\begin{lemma}[Recursion]\label{lem:recursion}
Let $0<W\subseteq T$ be split free of rank $\rho$ and $V=W\oplus W'$.
Embedding $\GL(W')$ by the identity on $W$, the restricted oracle hides
the stabilizer of the induced flag with members $U_i\cap W'$.
Remove repeated zero members. Each nonzero member is split free of rank
$\rank U_i-\rho$, and
\[
 U_i=W\oplus(U_i\cap W')\qquad(i<m).
\]
The new ambient rank is $n-\rho<n$, and a query costs one original oracle
call and $O(n^3)$ ring operations.
\end{lemma}
\begin{proof}
The decomposition follows from $W\subseteq U_i$.
Its second summand is a direct summand of $U_i$, hence of $V$ and $W'$.
An automorphism fixing $W$ pointwise and preserving $W'$ stabilizes $U_i$
exactly when its restriction stabilizes $U_i\cap W'$.
Strict rank gaps persist, except that the terminal member becomes zero
when $W=T$. In an adapted basis matrix $B$, the embedding is
$Y\mapsto B\operatorname{diag}(I_\rho,Y)B^{-1}$.
\end{proof}

\begin{theorem}[Flag reconstruction]\label{thm:qge3}
For $q\ge3$, Problem~\ref{prob:main} is solvable in bounded-error quantum
polynomial time. The algorithm also recovers the hidden flag.
\end{theorem}
\begin{proof}
For $t=1$ use \cite{DIKQS14}. Otherwise fix
$c=\lceil\log_4(2n/\epsilon)\rceil$ and $N=n(c+2)$.
Give each invocation of Proposition~\ref{prop:reduced},
Proposition~\ref{prop:branchA}, or Theorem~\ref{thm:hms-lift}
error at most $\epsilon/(2N)$.
A recursive call on $V'$ with oracle $f'$ proceeds as follows.
\begin{enumerate}[label=\arabic*.,leftmargin=*,itemsep=3pt]
\item Recover the reduced flag and its type. If it has one block, propose
the trivial flag and proceed to verification in step~4.
Otherwise let $d$ be its terminal rank.
\item If $d=1$, use Theorem~\ref{thm:hms-lift} to obtain a candidate $W$.
If $d\ge2$, try at most $c$ independent uniform unimodular $\lambda$,
using Proposition~\ref{prop:branchA} to form
$M_\lambda=\sum_X(X-I)V'$ from the returned generators.
In either case accept a candidate only if it is proper, nonzero, split
free, and passes $\mathcal L\subseteq H'$ from Lemma~\ref{lem:check}.
Reject failed draws; report failure if no candidate is accepted.
\item Put $\rho=\rank W$ and choose $V'=W\oplus W'$.
Recurse through Lemma~\ref{lem:recursion}.
Lift each nonzero member $U''$ of the returned flag to $W\oplus U''$,
and append $W$ exactly when $\rho=d$.
\item Check that the candidate is a split free flag of the type returned
in step~1. Compute its stabilizer generators and require
$f'(X)=f'(I)$ for every generator. Return the flag and generators if
all checks pass; otherwise return failure.
\end{enumerate}
All calls have fixed polynomial time limits; propagate failure returns
and cap the recursion at $n$ nodes.

\emph{Correctness with exact subroutine answers.}
Step~1 gives the true type. Step~2 certifies $0<W\subseteq T$, and a
transverse draw when $d\ge2$ passes with $W=T\cap P_\lambda$.
For nested split free modules, $\rho=d$ is equivalent to $W=T$.
Lemma~\ref{lem:recursion} therefore proves the assembly rule and the
strict decrease of ambient rank. The verification checks containment of
the candidate stabilizer in $H'$; the correct type gives equal orders
by Lemma~\ref{lem:conjugate}, hence equality.

\emph{Error and cost.}
In the ideal recursion with exact answers, each draw at a node with
$d\ge2$ is transverse with probability at least $3/4$.
Exhausting its $c$ draws has probability at most $4^{-c}$; across at
most $n$ nodes this contributes at most $\epsilon/2$.
Couple the actual recursion to this ideal execution until its first
incorrect subroutine answer. There are at most $N$ invocations, each
with a valid promise before that error, so discrepancy contributes
at most $\epsilon/2$.
The total failure probability is at most $\epsilon$.
The bounded recursion and all constituent bounds give
$\poly(n,\log|R|,\log(1/\epsilon))$ time on every execution.
The containment tests certify accepted $W$'s.
\end{proof}

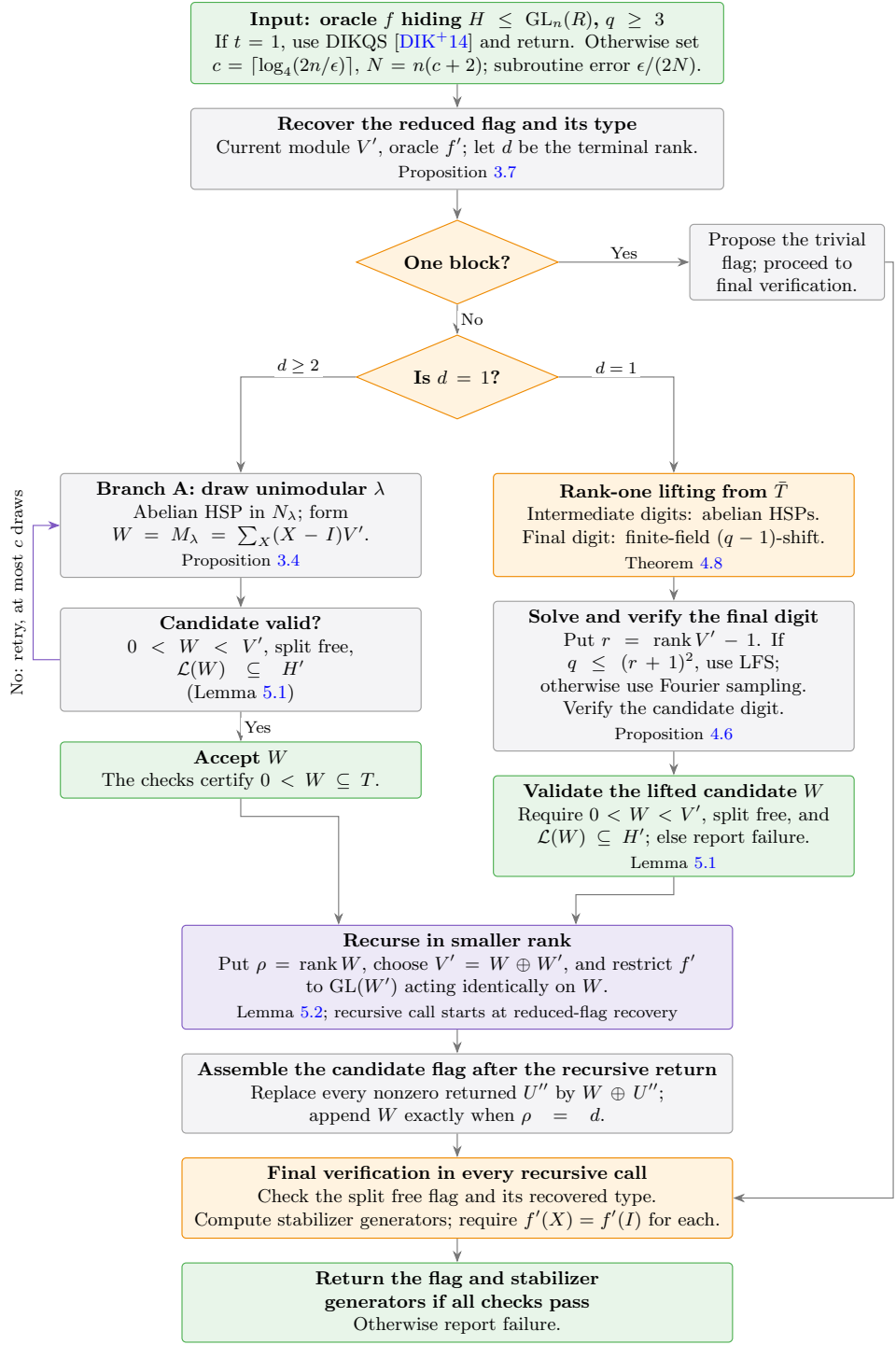
\begin{figure}[p]
\centering
\begin{adjustbox}{max width=\textwidth,max totalheight=.86\textheight,center}
\begin{tikzpicture}[node distance=4mm]
\node[fcio,text width=8.7cm] (in) at (0,0)
 {\textbf{Input: oracle $f$ hiding $H\le\GL_n(R)$, $q\ge3$}\\
  If $t=1$, use DIKQS~\cite{DIKQS14} and return. Otherwise set\\
  $c=\lceil\log_4(2n/\epsilon)\rceil$, $N=n(c+2)$;
  subroutine error $\epsilon/(2N)$.};
\node[fcstep,text width=8.7cm,below=of in] (rf)
 {\textbf{Recover the reduced flag and its type}\\
  Current module $V'$, oracle $f'$; let $d$ be the terminal rank.
  \fcref{Proposition~\ref{prop:reduced}}};
\node[fcdec,text width=2.4cm,below=5mm of rf] (one)
 {\textbf{One block?}};
\node[fcstep,text width=3.0cm,anchor=west] (triv) at ($(one.east)+(2.2,0)$)
 {Propose the trivial flag; proceed to final verification.};
\node[fcdec,text width=2.4cm,below=5mm of one] (d)
 {\textbf{Is $d=1$?}};
\node[fcstep,text width=5.8cm,anchor=north] (sample) at ($(d.south)+(-3.65,-.9)$)
 {\textbf{Branch A: draw unimodular $\lambda$}\\
  Abelian HSP in $N_\lambda$; form\\
  $W=M_\lambda=\sum_X(X-I)V'$.
  \fcref{Proposition~\ref{prop:branchA}}};
\node[fcstep,text width=5.8cm,below=5mm of sample] (test)
 {\textbf{Candidate valid?}\\
  $0<W<V'$, split free,\\
  $\mathcal L(W)\subseteq H'$\\
  (Lemma~\ref{lem:check})};
\node[fcio,text width=5.8cm,below=5mm of test] (accept)
 {\textbf{Accept $W$}\\
  The checks certify $0<W\subseteq T$.};
\node[fckey,text width=5.8cm,anchor=north] (rank) at ($(d.south)+(3.65,-.9)$)
 {\textbf{Rank-one lifting from $\bar T$}\\
  Intermediate digits: abelian HSPs.\\
  Final digit: finite-field $(q-1)$-shift.
  \fcref{Theorem~\ref{thm:hms-lift}}};
\node[fcstep,text width=5.8cm,below=of rank] (shift)
 {\textbf{Solve and verify the final digit}\\
  Put $r=\rank V'-1$. If $q\le(r+1)^2$, use LFS;\\
  otherwise use Fourier sampling.\\
  Verify the candidate digit.
  \fcref{Proposition~\ref{prop:finite-field-shift}}};
\node[fcio,text width=5.8cm,below=of shift] (rtest)
 {\textbf{Validate the lifted candidate $W$}\\
  Require $0<W<V'$, split free, and\\
  $\mathcal L(W)\subseteq H'$; else report failure.
  \fcref{Lemma~\ref{lem:check}}};
\node[fit=(accept)(rtest),inner sep=0pt] (branches) {};
\node[fcbox,fill=fcPurple,draw=fcVioletE,text width=9.0cm,anchor=north] (rec) at ($(0,0 |- branches.south)+(0,-.8)$)
 {\textbf{Recurse in smaller rank}\\
  Put $\rho=\rank W$, choose $V'=W\oplus W'$, and restrict $f'$\\
  to $\GL(W')$ acting identically on $W$.
  \fcref{Lemma~\ref{lem:recursion}; recursive call starts at reduced-flag recovery}};
\node[fcstep,text width=9.0cm,below=of rec] (assemble)
 {\textbf{Assemble the candidate flag after the recursive return}\\
  Replace every nonzero returned $U''$ by $W\oplus U''$;\\
  append $W$ exactly when $\rho=d$.};
\node[fckey,text width=9.0cm,below=of assemble] (verify)
 {\textbf{Final verification in every recursive call}\\
  Check the split free flag and its recovered type.\\
  Compute stabilizer generators; require $f'(X)=f'(I)$ for each.};
\node[fcio,text width=9.0cm,below=of verify] (out)
 {\textbf{Return the flag and stabilizer generators if all checks pass}\\
  Otherwise report failure.};
\draw[fcflow] (in)--(rf);
\draw[fcflow] (rf)--(one);
\draw[fcflow] (one)--(triv) node[fclab,midway,above]{Yes};
\draw[fcflow] (one)--(d) node[fclab,midway,right]{No};
\draw[fcflow] (d.west) -| (sample.north) node[fclab,pos=.25,above]{$d\ge2$};
\draw[fcflow] (d.east) -| (rank.north) node[fclab,pos=.25,above]{$d=1$};
\draw[fcflow] (sample)--(test);
\draw[fcflow] (test)--(accept) node[fclab,midway,right]{Yes};
\draw[fcloop] (test.west)--(-7.15,0 |- test.west) |- (sample.west);
\node[fclab,rotate=90] at ($(sample.west)!0.5!(test.west)+(-.7,0)$) {No: retry, at most $c$ draws};
\draw[fcflow] (rank)--(shift);
\draw[fcflow] (shift)--(rtest);
\draw[fcflow] (accept.south) -- ++(0,-.3) -| ([xshift=-2cm]rec.north);
\draw[fcflow] (rtest.south) -- ++(0,-.3) -| ([xshift=2cm]rec.north);
\draw[fcflow] (rec)--(assemble);
\draw[fcflow] (assemble)--(verify);
\draw[fcflow] (verify)--(out);
\draw[fcflow] (triv.east) -- (7.35,0 |- triv.east) |- (verify.east);
\end{tikzpicture}
\end{adjustbox}
\caption{Full reconstruction for $q\ge3$ (Theorem~\ref{thm:qge3}).
The one-block case also undergoes final verification. Exhausted draws,
failed checks, timeouts and recursive failures return failure; all loops
are capped and there are at most $n$ recursive nodes. The overall success
probability is at least $1-\epsilon$.}
\label{fig:qge3-flow}
\end{figure}
\clearpage

\section{Subgroup lifting for binary residue fields}\label{sec:q2}
Assume $k=\F_2$ and $t\ge2$. Let $P$ be the standard parabolic (Definition~\ref{def:split}) of the
recovered type, $\mathfrak l\le\Mat_n(k)$ its block lower triangular
algebra, and $\mathfrak u$ the strictly upper-block subspace, so
$\Mat_n(k)=\mathfrak l\oplus\mathfrak u$.
Write $\operatorname{pr}_{\mathfrak u}$ for the projection, and put
$\varepsilon=\pi^{t-1}$ and
\[
 \iota(Y)=I+\varepsilon\widehat Y\qquad(Y\in\Mat_n(k) \text{ and }\widehat Y=\sigma(Y)\in \Mat_n(R)).
\]
This does not depend on the lift $\widehat Y$, since
$\varepsilon\m=0$; also $\varepsilon^2=0$.
We recover the conjugate subgroup rather than a uniquely specified flag.
In this section $\varepsilon$ always denotes $\pi^{t-1}$, not the error parameter $\epsilon$. The map $\iota$ is the identification $\Mat_n(k)\cong K_{t-1}$ noted after Definition~\ref{def:congruence}. Under it, $P\cap K_{t-1}$ corresponds to $\mathfrak l$.

\begin{lemma}[Congruence commutators]\label{lem:q2-comm}
For $1\le a,b\le t$, $A\in\pi^a\Mat_n(R)$ and $B\in\pi^b\Mat_n(R)$,
one has $[K_a,K_b]\subseteq K_{a+b}$.
If $a+b\ge t-1$, then
\[
 [I+A,I+B]=I+AB-BA.
\]
The coefficient depends only on $A\bmod\pi^{t-b}$ and
$B\bmod\pi^{t-a}$.
\end{lemma}
\begin{proof}
The exact identity
\[
 [I+A,I+B]
 =I+(AB-BA)(I+A)^{-1}(I+B)^{-1}
\]
gives the inclusion. When $a+b\ge t-1$, multiplication of $AB-BA$
by either inverse leaves it unchanged because $\pi^{t-1}\m=0$.
The two dependence claims follow by multiplying the possible differences.
\end{proof}
\noindent This lemma is what makes digit extraction possible. Suppose $g\in K_a$ is known only modulo $\pi^{a+1}$, and take $B\in\pi^{t-1-a}\Mat_n(R)$. Then $[g,I+B]$ already lies in the socle layer $K_{t-1}$, and it depends only on the leading digit of $g-I$.

\subsection{Standardization and the reference oracle}

\begin{lemma}[Standardization after reduction]\label{lem:q2-standard}
Let $\Fl$ be a split free flag whose reduction modulo $\m$ is the
standard flag $\Fl_0$ stabilized by $P$. Then
\[
 \Fl=g\Fl_0
\]
for some $g\in K_1$. Consequently
\[
 G_{\{\Fl\}}=gPg^{-1},\qquad g\in K_1.
\]
\end{lemma}

\begin{proof}
Choose a basis of $V$ adapted to $\Fl$ whose reduction is the standard
basis of $\bar V$. Such a basis is obtained by lifting, successively
along the flag, bases of the quotients $\bar U_i/\bar U_{i+1}$; the
basis-extension property in Lemma~\ref{lem:splitting} and Nakayama's
lemma show that the lifts form an $R$-basis. If $g$ is the matrix whose
columns are these vectors, then $g\equiv I\pmod\m$, hence $g\in K_1$,
and $g\Fl_0=\Fl$.
\end{proof}

\begin{lemma}[Socle reference oracle]\label{lem:q2-reference}
Suppose
\[
 H=gPg^{-1},\qquad g\in K_1.
\]
Then
\[
 H\cap K_{t-1}=P\cap K_{t-1}.
\]
Moreover
\[
 A_{\mathfrak u}:=\{\iota(Y):Y\in\mathfrak u\}
\]
is an elementary abelian subgroup isomorphic to $(\mathfrak u,+)$, and
\[
 F:\mathfrak u\longrightarrow S,\qquad F(Y)=f(\iota(Y)),
\]
is injective.
\end{lemma}
\noindent Socle elements commute with $K_1$, so conjugation by the unknown $g$ is invisible on $K_{t-1}$. There the oracle sees only the known $P$, and $F$ labels the cosets of $P\cap K_{t-1}\leftrightarrow\mathfrak l$ by their $\mathfrak u$-component. This gives a reference copy of the oracle against which shifts can be measured.

\begin{proof}
The inclusion $[K_1,K_{t-1}]\subseteq K_t=\{I\}$ gives
$H\cap K_{t-1}=P\cap K_{t-1}$.
The identity $\iota(Y)\iota(Z)=\iota(Y+Z)$ identifies $A_{\mathfrak u}$
with $(\mathfrak u,+)$.
If $f(\iota(Y))=f(\iota(Z))$, then
$\iota(Z-Y)\in P\cap K_{t-1}$, hence $Z-Y\in\mathfrak l$.
For $Y,Z\in\mathfrak u$ this forces $Y=Z$.
\end{proof}

\begin{lemma}[Two-copy socle shift]\label{lem:q2-shift}
Assume $H=gPg^{-1}$ with $g\in K_1$, and let $w\in P$ satisfy
\[
 gwg^{-1}w^{-1}=\iota(C)
\]
for some $C\in M_n(k)$. Put
\[
 s=\operatorname{pr}_{\mathfrak u}(C).
\]
Then the function
\[
 \Phi_w:\mathfrak u\times\F_2\longrightarrow S
\]
defined by
\[
 \Phi_w(Y,0)=f(\iota(Y)),\qquad
 \Phi_w(Y,1)=f(\iota(Y)w)
\]
hides the subgroup
\[
 L_s=\{(0,0),(s,1)\}
 \le \mathfrak u\times\F_2.
\]
Thus $s$ is recoverable by an abelian HSP in quantum time
$\poly(n,\log|R|)$.
\end{lemma}
\noindent Here $w$ is a known element of the reference parabolic $P$, and conjugation by the unknown $g$ moves it by the socle element $\iota(C)$. The oracle on $\iota(Y)w$ is therefore the reference oracle $F$ shifted by $s$, the $\mathfrak u$-part of $C$. In characteristic two, shifting by $s$ while flipping the second coordinate is an involution, so $\{(0,0),(s,1)\}$ is a subgroup. This is the ``two-copy'' structure, and it replaces the $q-1$ shifted copies available when $q\ge3$.

\begin{proof}
Put $c=\iota(C)$, so $cw=gwg^{-1}\in H$.
Right multiplication by $(cw)^{-1}$ gives
\[
 \Phi_w(Y,1)=f(\iota(Y)w(cw)^{-1})=f(\iota(Y-C)).
\]
Write $C=C_{\mathfrak l}+s$. Since
$\iota(-C_{\mathfrak l})\in P\cap K_{t-1}\subseteq H$,
we obtain $\Phi_w(Y,1)=F(Y-s)$ and $\Phi_w(Y,0)=F(Y)$.
Injectivity of $F$ and characteristic two show that collisions are
exactly differences in $\{(0,0),(s,1)\}$.
\end{proof}

The HSP here is on the parameter group $\mathfrak u\times\F_2$.
The query matrices $\iota(Y)w^b$ need not form a subgroup; their
multiplication is not used in the reduction.

\subsection{Intermediate layers}

\begin{proposition}[One intermediate lifting layer]\label{prop:q2-layer}
Suppose
\[
 H_j=g_jPg_j^{-1},\qquad g_j\in K_j,
\]
where $1\le j<t-1$. Using $n$ abelian HSP instances one can compute an
explicit $c_j\in K_j$ such that
\[
 c_jH_jc_j^{-1}=g_{j+1}Pg_{j+1}^{-1}
\]
for some $g_{j+1}\in K_{j+1}$.
\end{proposition}

\begin{proof}
Write $g_j\equiv I+\pi^j\widehat X\pmod{\pi^{j+1}}$ and
$x=\operatorname{pr}_{\mathfrak u}(X)$.
For $a=1,\ldots,n$, use the diagonal element
$w_a=I+\pi^{t-1-j}E_{aa}\in P$.
Since $g_j\in K_j$ and $w_a\in K_{t-1-j}$, Lemma~\ref{lem:q2-comm} applies with exponents summing to $t-1$. The commutator lies in the socle and depends only on the leading digit $X$ of $g_j$:
\[
 [g_j,w_a]=\iota([X,E_{aa}]).
\]
The two-copy HSP of Lemma~\ref{lem:q2-shift} therefore returns
$s_a=\operatorname{pr}_{\mathfrak u}([X,E_{aa}])
=\operatorname{pr}_{\mathfrak u}([x,E_{aa}])$.
The lower-block component drops out because $\mathfrak l$ is an algebra.
For every upper-block entry in column $a$,
$[x,E_{aa}]_{ia}=x_{ia}$, so these shifts determine $x$.

Lift $x$ with support in the upper blocks and put $c_j=I-\pi^j\widehat x$.
The leading coefficient of $c_jg_j$ lies in $\mathfrak l$.
Thus for some $p_j\in P\cap K_j$,
$g_{j+1}=c_jg_jp_j^{-1}\in K_{j+1}$, and
$c_jH_jc_j^{-1}=g_{j+1}Pg_{j+1}^{-1}$.
Only $c_j$ needs to be computed.
\end{proof}

Updating the oracle by $f_{j+1}(X)=f_j(c_j^{-1}Xc_j)$ removes one layer.
Iterate for $j=1,\ldots,t-2$.

\subsection{The final layer}
Now $H_{t-1}=gPg^{-1}$ with $g\in K_{t-1}$.
Absorb the $\mathfrak l$-component into $P$ and write $g=\iota(X)$
for some $X\in\mathfrak u$. Let $\bar P=P(k)$.

\begin{proposition}[Final layer]\label{prop:q2-final}
The subgroup $H_{t-1}$ can be recovered in quantum time
$\poly(n,\log|R|)$.
\end{proposition}
\noindent At the last layer the conjugator $g=\iota(X)$ lies in the socle itself, so its commutator with every $w\in P$ is a socle element, not only with diagonal probes. Each generator $u_i$ of $\bar P$ gives one linear equation for $X$ over $\F_2$. Here $\rho$ is the action of $\bar P$ on $\mathfrak u\cong\Mat_n(k)/\mathfrak l$ induced by conjugation. The solution set is $X$ plus the directions $D$ with $\iota(D)$ normalizing $P$, and these do not change $H_{t-1}$.

\begin{proof}
Choose generators $u_1,\ldots,u_\nu$ of $\bar P$, using the admissible
elementary transvections over $\F_2$; $\nu\le n(n-1)$.
Lift them to $w_i\in P$. Since $\varepsilon^2=0$,
\[
 [g,w_i]=\iota(X-u_iXu_i^{-1}),
\]
so Lemma~\ref{lem:q2-shift} yields
$s_i=\operatorname{pr}_{\mathfrak u}(X-u_iXu_i^{-1})$.

Conjugation preserves $\mathfrak l$, hence induces the representation
\[
 \rho(u)Y=\operatorname{pr}_{\mathfrak u}(uYu^{-1})
 \quad\text{on }\mathfrak u\cong\Mat_n(k)/\mathfrak l.
\]
Solve $(I-\rho(u_i))X_0=s_i$ over $\F_2$.
The true $X$ proves consistency. For any solution, $D=X-X_0$ is fixed
by the generators and therefore by all of $\bar P$.
Equivalently, $D-uDu^{-1}\in\mathfrak l$ for every $u\in\bar P$.
For $p\in P$ this gives
\[
 [\iota(D),p]
 =I+\varepsilon(\widehat D-p\widehat Dp^{-1})\in P.
\]
Thus $\iota(D)P\iota(D)^{-1}\subseteq P$, and finiteness gives equality.
Conversely, if $\iota(D)$ normalizes $P$, these commutators belong to
$P\cap K_{t-1}$; their reduced coefficients are in $\mathfrak l$,
so $\rho(u)D=D$ for all $u$.
The homogeneous solutions are therefore exactly the socle normalizer
directions.

Finally, $\iota(X)=\iota(X_0)\iota(D)$, and hence
$H_{t-1}=\iota(X_0)P\iota(X_0)^{-1}$.
Conjugating generators of $P$ gives the required output.
\end{proof}

\begin{theorem}[Binary residue field]\label{thm:q2}
Let $R$ be a finite chain ring with residue field $R/\m=\F_2$, presented as in
Section~\ref{sec:model}. Then Problem~\ref{prob:main} is solvable in quantum
time $\poly(n,\log|R|)$, in the sense of Convention~\ref{conv:error}.
\end{theorem}

\begin{proof}
For $t=1$ use \cite{DIKQS14}. Otherwise set
\[
 N_0=1+n(t-2)+n(n-1)
\]
and give each of the following randomized calls error at most
$\epsilon/N_0$. First recover $\bar\Fl$ by
Proposition~\ref{prop:reduced}. Let $\bar b$ have columns forming an
adapted basis of $\bar\Fl$, and lift it to $b\in\GL_n(R)$.
Then $b^{-1}Hb=g_1Pg_1^{-1}$ for some $g_1\in K_1$ by
Lemma~\ref{lem:q2-standard}.

Apply Proposition~\ref{prop:q2-layer} for $j=1,\ldots,t-2$,
then Proposition~\ref{prop:q2-final}.
The resulting subgroup is $CHC^{-1}$, where the known product is
\[
 C=c_{t-2}\cdots c_1b^{-1}.
\]
Conjugate its generators by $C^{-1}$ to obtain generators of $H$.

Before the first erroneous answer, every call has a valid promise.
There are at most $N_0$ calls, so the first-error union bound gives
failure probability at most $\epsilon$.
Use the prescribed time caps and return failure on malformed subspaces,
inconsistent linear systems, or failed subroutines.
Each HSP has an elementary abelian parameter group of dimension at most
$n^2+1$ over $\F_2$; the remaining work is polynomial linear algebra
and matrix arithmetic. Since $t\le\log_2|R|$, the claimed bounded
running time follows. For the one-block type, $\mathfrak u=0$ and the
same construction simply returns $G$.
\end{proof}

\begin{remark}[Even residue fields]\label{rem:q2-extension}
The same lifting construction works over $k=\F_{2^f}$.
Its two-copy graph is still a subgroup of $\mathfrak u\times\F_2$.
Generate $\bar P$ by admissible $I+b_rE_{ij}$ for an $\F_2$-basis
$\{b_r\}$ of $k$, together with diagonal generators of $k^\times$;
there are $O(n^2\log q)$ such generators.
Expand the parameter groups and final linear system over $\F_2$.
All costs remain polynomial in $n,\log|R|$.
In odd characteristic a nonzero shift does not give a subgroup of
$\mathfrak u\times\F_2$, so this two-copy reduction is unavailable.
\end{remark}

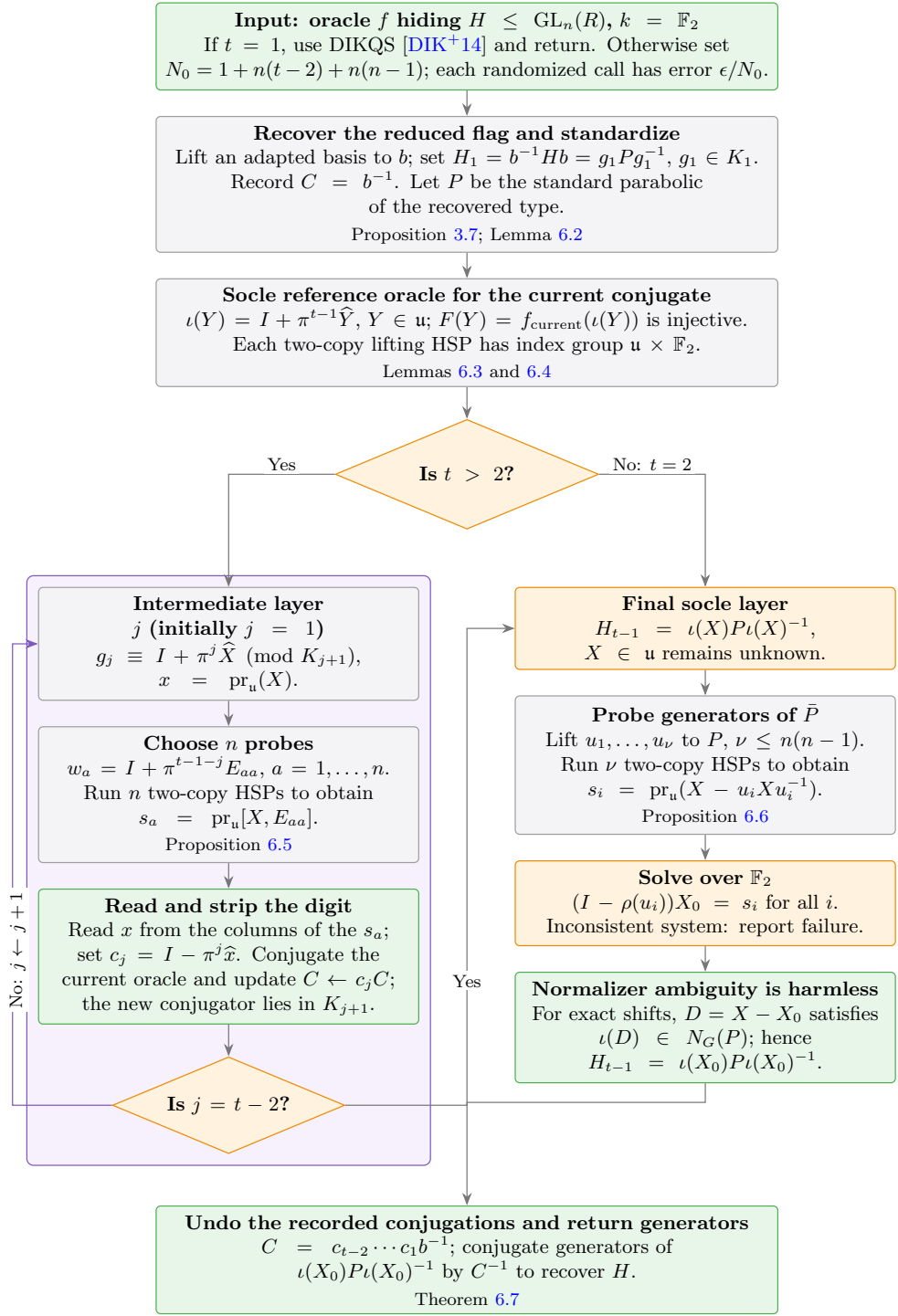
\begin{figure}[p]
\centering
\begin{adjustbox}{max width=\textwidth,max totalheight=.86\textheight,center}
\begin{tikzpicture}[node distance=4mm]
\node[fcio,text width=9.5cm] (in) at (0,0)
 {\textbf{Input: oracle $f$ hiding $H\le\GL_n(R)$, $k=\F_2$}\\
  If $t=1$, use DIKQS~\cite{DIKQS14} and return. Otherwise set\\
  $N_0=1+n(t-2)+n(n-1)$; each randomized call has error $\epsilon/N_0$.};
\node[fcstep,text width=9.5cm,below=of in] (reduced)
 {\textbf{Recover the reduced flag and standardize}\\
  Lift an adapted basis to $b$; set $H_1=b^{-1}Hb=g_1Pg_1^{-1}$, $g_1\in K_1$.\\
  Record $C=b^{-1}$. Let $P$ be the standard parabolic\\
  of the recovered type.
  \fcref{Proposition~\ref{prop:reduced}; Lemma~\ref{lem:q2-standard}}};
\node[fcstep,text width=9.5cm,below=of reduced] (ref)
 {\textbf{Socle reference oracle for the current conjugate}\\
  $\iota(Y)=I+\pi^{t-1}\widehat Y$, $Y\in\mathfrak u$;
  $F(Y)=f_{\rm current}(\iota(Y))$ is injective.\\
  Each two-copy lifting HSP has index group $\mathfrak u\times\F_2$.
  \fcref{Lemmas~\ref{lem:q2-reference} and~\ref{lem:q2-shift}}};
\node[fcdec,text width=3.1cm,below=5mm of ref] (layers)
 {\textbf{Is $t>2$?}};
\node[fcstep,text width=5.7cm,anchor=north] (layer) at ($(layers.south)+(-3.75,-.9)$)
 {\textbf{Intermediate layer $j$ (initially $j=1$)}\\
  $g_j\equiv I+\pi^j\widehat X\pmod{K_{j+1}}$,\\
  $x=\operatorname{pr}_{\mathfrak u}(X)$.};
\node[fcstep,text width=5.7cm,below=of layer] (probes)
 {\textbf{Choose $n$ probes}\\
  $w_a=I+\pi^{t-1-j}E_{aa}$, $a=1,\ldots,n$.\\
  Run $n$ two-copy HSPs to obtain\\
  $s_a=\operatorname{pr}_{\mathfrak u}[X,E_{aa}]$.
  \fcref{Proposition~\ref{prop:q2-layer}}};
\node[fcio,text width=5.7cm,below=of probes] (digit)
 {\textbf{Read and strip the digit}\\
  Read $x$ from the columns of the $s_a$;\\
  set $c_j=I-\pi^j\widehat x$. Conjugate the\\
  current oracle and update $C\leftarrow c_jC$;\\
  the new conjugator lies in $K_{j+1}$.};
\node[fcdec,text width=2.5cm,below=5mm of digit] (last)
 {\textbf{Is $j=t-2$?}};
\node[fckey,text width=5.7cm,anchor=north] (final) at ($(layers.south)+(3.75,-.9)$)
 {\textbf{Final socle layer}\\
  $H_{t-1}=\iota(X)P\iota(X)^{-1}$,\\
  $X\in\mathfrak u$ remains unknown.};
\node[fcstep,text width=5.7cm,below=of final] (gens)
 {\textbf{Probe generators of $\bar P$}\\
  Lift $u_1,\ldots,u_\nu$ to $P$, $\nu\le n(n-1)$.\\
  Run $\nu$ two-copy HSPs to obtain\\
  $s_i=\operatorname{pr}_{\mathfrak u}(X-u_iXu_i^{-1})$.
  \fcref{Proposition~\ref{prop:q2-final}}};
\node[fckey,text width=5.7cm,below=of gens] (solve)
 {\textbf{Solve over $\F_2$}\\
  $(I-\rho(u_i))X_0=s_i$ for all $i$.\\
  Inconsistent system: report failure.};
\node[fcio,text width=5.7cm,below=of solve] (normal)
 {\textbf{Normalizer ambiguity is harmless}\\
  For exact shifts, $D=X-X_0$ satisfies\\
  $\iota(D)\in N_G(P)$; hence\\
  $H_{t-1}=\iota(X_0)P\iota(X_0)^{-1}$.};
\node[fit=(last)(normal),inner sep=0pt] (branches) {};
\node[fcio,text width=9.5cm,anchor=north] (out) at ($(0,0 |- branches.south)+(0,-.8)$)
 {\textbf{Undo the recorded conjugations and return generators}\\
  $C=c_{t-2}\cdots c_1b^{-1}$; conjugate generators of\\
  $\iota(X_0)P\iota(X_0)^{-1}$ by $C^{-1}$ to recover $H$.
  \fcref{Theorem~\ref{thm:q2}}};
\begin{scope}[on background layer]
\node[draw=fcVioletE,fill=fcViolet,rounded corners=3pt,line width=.5pt,
 fit=(layer)(probes)(digit)(last),inner sep=5pt] {};
\end{scope}
\draw[fcflow] (in)--(reduced);
\draw[fcflow] (reduced)--(ref);
\draw[fcflow] (ref)--(layers);
\draw[fcflow] (layers.west) -| (layer.north) node[fclab,pos=.25,above]{Yes};
\draw[fcflow] (layers.east) -| (final.north) node[fclab,pos=.25,above]{No: $t=2$};
\draw[fcflow] (layer)--(probes);
\draw[fcflow] (probes)--(digit);
\draw[fcflow] (digit)--(last);
\draw[fcloop] (last.west) -- (-7.15,0 |- last.west) |- (layer.west);
\node[fclab,rotate=90] at ($(probes.west)!0.5!(last.west)+(-.9,0)$) {No: $j\leftarrow j+1$};
\draw[fcflow] (last.east) -- (0,0 |- last.east) |- (final.west)
 node[fclab,pos=.12,above]{Yes};
\draw[fcflow] (final)--(gens);
\draw[fcflow] (gens)--(solve);
\draw[fcflow] (solve)--(normal);
\draw[fcflow] (normal.south) -- ++(0,-.3) -| (out.north);
\end{tikzpicture}
\end{adjustbox}
\caption{Subgroup reconstruction for $q=2$ (Theorem~\ref{thm:q2}).
The intermediate loop is skipped when $t=2$. All calls have prescribed
time caps; malformed outputs and failed subroutines return failure.
With probability at least $1-\epsilon$, the returned generators define
$H$; a unique hidden flag is not required. For the one-block type,
$\mathfrak u=0$ and the construction returns $\GL_n(R)$.}
\label{fig:q2-flow}
\end{figure}
\clearpage

\section{Arithmetic consequences}\label{sec:applications}
\begin{proof}[Proof of Theorem~\ref{thm:main}]
Apply Theorem~\ref{thm:qge3} when $q\ge3$ and
Theorem~\ref{thm:q2} when $q=2$, with the common input and error model.
\end{proof}

\begin{corollary}\label{cor:ZN}
For every integer $N\ge2$, the parabolic HSP in
$\GL_n(\Z/N\Z)$ is solvable in quantum time $\poly(n,\log N)$.
\end{corollary}

\begin{proof}
Factor $N$ using Shor's algorithm \cite{Sho97} and apply CRT:
$\Z/N\Z\cong\prod_{p^a\parallel N}\Z/p^a\Z$.
A split free flag decomposes componentwise, as does its stabilizer.
The embedding of one matrix-group factor with identities in the others
therefore restricts the oracle to a component parabolic.
Apply Theorem~\ref{thm:main} to each component and combine generators
by CRT. This includes all powers of two.
There are at most $\log_2N$ factors and their logarithmic sizes sum to
$\log N$. Assign error $\epsilon/2$ to integer factorization and split
the remaining budget among the component calls.
\end{proof}

\begin{definition}[Number-field input]\label{def:LK}
A number field $K$ is supplied by a monic irreducible $g\in\Z[x]$,
an integral basis of its maximal order $\OK$ expressed in the power basis
of $K=\Q[x]/(g)$, and its multiplication table.
Let $L_K$ be the total bit length of this data.
\end{definition}
The maximal-order basis is part of the input; computing it is not assumed
free. Arithmetic modulo $N$ is polynomial in $L_K,\log N$.

\begin{corollary}\label{cor:numberfield}
Let $K$ be a number field given in standard form and let $N\ge2$. Then
the parabolic HSP in
\[
 \GL_n(\OK/N\OK)
\]
is solvable in quantum time $\poly(n,L_K,\log N,\log(1/\epsilon))$ with
probability at least $1-\epsilon$.
\end{corollary}

\begin{proof}
After factoring $N$, write
$p\OK=\prod_{\p\mid p}\p^{e_\p}$ and set $a_\p=e_\p v_p(N)$.
Then
\[
 \OK/N\OK\cong\prod_{\p\mid N}\OK/\p^{a_\p},
\]
and each factor is a finite chain ring.
Appendix~\ref{app:numberfield} constructs the required effective
presentations in bounded randomized classical polynomial time once
the factorization of $N$ is available.
Restrict the oracle to each factor as in Corollary~\ref{cor:ZN},
apply Theorem~\ref{thm:main}, and recombine.
The logarithmic factor sizes sum to $[K:\Q]\log N$.
Splitting the error budget between factorization, preprocessing, and
the component calls preserves the stated bound.
\end{proof}

\section*{Declaration on the use of AI tools}
The author used Claude Opus 5.5 to assist with \LaTeX{} formatting, checking grammatical errors, and exploring proof strategies. All mathematical results, proofs and claims were verified by the author, who takes full responsibility for the content of this paper.

\section*{Acknowledgment}
The author is supported by EPSRC through grant number EP/V011324/1. The author also thanks Gábor Ivanyos for reading an earlier version of this manuscript and for
helpful comments.
\bibliographystyle{alpha}
\bibliography{reference}

\clearpage
\appendix
\section{Supporting algebra and effective operations}\label{app:algebra}

\subsection{Split submodules and faithfulness}
\begin{proof}[Proof of Lemma~\ref{lem:splitting}]
Given generators with independent reductions, extend those reductions to
a basis of $k^n$ and lift the additional vectors. The resulting square
matrix has unit determinant, so it is an $R$-basis of $V$; this proves
(b)$\Rightarrow$(a).
If $V=U\oplus C$, then $U\cap\pi^jV=\pi^jU$, proving purity and
$\bar U\cong U/\m U$; hence (a)$\Rightarrow$(c).
Under (c), lifts of a basis of $\bar U\cong U/\m U$ generate $U$
by Nakayama, proving (b).

A complement of a free direct summand is a finitely generated projective
module over the local ring $R$, hence free \cite{AM69,McDonald74}.
If $U'\le U$ is split free with $V=U'\oplus D$, then
$U=U'\oplus(U\cap D)$. Finally $\rank U=\dim_k\bar U$ implies that
a nonzero split free module cannot lie in $\m V$.
\end{proof}

\begin{lemma}[Invariant subspaces of a parabolic over a field]
\label{lem:field-invariant}
Let $q\ge3$, let $\bar\Fl:k^n=\bar U_0>\bar U_1>\dots>\bar U_m=0$ be a flag of
subspaces of $k^n$ and let $\bar P\le\GL_n(k)$ be its stabilizer. Then the
$\bar P$-invariant subspaces of $k^n$ are exactly the members
$\bar U_0,\dots,\bar U_m$.
\end{lemma}

\begin{proof}
Use the block decomposition $k^n=\bigoplus_{j=1}^mB_j$ with
$\bar U_i=\bigoplus_{j>i}B_j$.
Choose $\alpha\in k^\times\setminus\{1\}$.
Scaling only block $j$ by $\alpha$ shows that any invariant subspace
$\bar W$ contains each block component of each of its vectors.
Thus $\bar W=\bigoplus_j(\bar W\cap B_j)$.
The block group $\GL(B_j)$ is transitive on nonzero vectors, so every
intersection is $0$ or $B_j$.
Lower-block transvections force the selected blocks to be a terminal
segment, giving exactly the stated flag members.
\end{proof}

\begin{proof}[Proof of Theorem~\ref{thm:faithful}]
The idea is as follows. Reducing modulo $\m$ pins $\bar W$ down to a reduced member $\bar U_i$. A block scalar $X_\alpha\in H$ with $\alpha-1$ a unit, available exactly because $q\ge3$, then splits $W$ along $U_i\oplus C_i$, and Nakayama's lemma finishes the argument.

Write $H=G_{\{\Fl\}}$ for $\Fl:V=U_0>\dots>U_m=0$, and let $W$ be a member of $\Fl'$, so $W$ is split free and $H$-invariant. We show $W=U_i$ for some $i$.
 
Fix an adapted basis for $\Fl$ and let $C_i$ be the span of the basis
vectors outside $U_i$, so $V=U_i\oplus C_i$ and $U_j=U_i\oplus(U_j\cap
C_i)$ for $j<i$. Reduction modulo $\m$ maps $H$ onto the parabolic
subgroup $\bar P$ of $\GL_n(k)$ stabilizing $\bar\Fl$ (block triangular matrices
lift to block triangular matrices), whose invariant subspaces are exactly
the $\bar U_i$ by Lemma~\ref{lem:field-invariant}. Hence $\bar W=\bar U_i$ for some $i$.
 
For $\alpha\in R^\times$ let $X_\alpha$ act as $\alpha$ on $U_i$ and as
the identity on $C_i$. Then $X_\alpha$ preserves every $U_j$, so $X_\alpha\in H$. For $w=u+c\in W$ with $u\in U_i$, $c\in C_i$ we get
$X_\alpha w-w=(\alpha-1)u\in W$. By \eqref{eq:units} we may choose
$\alpha$ with $\alpha-1\in R^\times$, whence $u\in W$ and $c\in W$. Thus
$W=(W\cap U_i)\oplus(W\cap C_i)$, and $W\cap C_i$ is a direct summand of
$W$, hence split free in $V$ by Lemma~\ref{lem:splitting}. But
$\overline{W\cap C_i}\subseteq\bar W\cap\bar C_i=\bar U_i\cap\bar C_i=0$,
so $W\cap C_i\subseteq\m V$, which forces $W\cap C_i=0$ by the last
sentence of Lemma~\ref{lem:splitting}. Therefore $W\subseteq U_i$, and
$\bar W=\bar U_i$ gives $U_i=W+\m U_i$, so $W=U_i$ by Lemma~\ref{lem:nak}.
\end{proof}

\begin{proposition}[Failure for $q=2$]\label{prop:unfaithful}
Let $q=2$, $t\ge2$, $n=2$, and let $S=Rv$ be a free cyclic submodule of
$V=R^2$. Then for a free cyclic $S'=Rv'$,
\[
  G_{\{S'\}}=G_{\{S\}}\quad\Longleftrightarrow\quad
  v'\equiv uv \pmod{\pi^{t-1}V}\ \text{for some }u\in R^\times ,
\]
so the fibre of $\Fl\mapsto G_{\{\Fl\}}$ through $S$ has exactly $q=2$
elements. In particular the oracle of Problem~\ref{prob:main} does not
determine the last $\pi$-adic digit of $S$.
\end{proposition}

\begin{proof}
Choose $S=Re_1$; its stabilizer consists of
$\left(\begin{smallmatrix}a&b\\0&d\end{smallmatrix}\right)$ with
$a,d\in R^\times$.
If a unimodular $(x,y)$ spans an invariant line and $y$ is a unit,
varying $b$ gives a contradiction. Normalize therefore to $(1,y)$
with $y\in\m$. Invariance is equivalent to
\[
 (d-a)y-by^2=0\qquad(a,d\in R^\times,\ b\in R).
\]
Since $R^\times=1+\m$, the differences $d-a$ run through $\m$.
Thus $\m y=0$, or $y\in\pi^{t-1}R$.
Conversely this condition also gives $y^2=0$ for $t\ge2$, so it is
sufficient. There are exactly $q=2$ choices of $y$.
We have proved containment of stabilizers; both stabilize split lines,
so they are conjugate and have equal order. Containment is therefore
equality.
\end{proof}

\begin{example}\label{ex:z4}
Over $\Z/4\Z$, the lines $R(1,0)$ and $R(1,2)$ are distinct but have
the same stabilizer. Indeed an upper triangular invertible matrix sends
$(1,2)$ to $(a+2b,2d)=(a+2b)(1,2)$, since $a,d$ are odd.
\end{example}

\subsection{Unit generators and module algorithms}\label{app:effective}
Let $b_1,\ldots,b_f$ be an $\F_p$-basis of $k$, and let $\gamma$ lift
a generator of $k^\times$. Then
\[
 \gamma,\qquad 1+\pi^i\sigma(b_r)
 \quad(1\le i<t,\ 1\le r\le f)
\]
generate $R^\times$. Their images generate the residue units and every
quotient $(1+\m^i)/(1+\m^{i+1})\cong(k,+)$; induction through the
filtration proves the claim. There are $1+(t-1)f=O(\log|R|)$ generators.

A certified primitive residue element can be found in quantum polynomial
time by factoring $q-1$ with Shor's algorithm and testing orders.
For randomized classical preprocessing, use instead $h$ independent
uniform elements of $k^\times$. The cyclic group has one maximal subgroup
for each prime $\ell\mid q-1$, so failure to generate has probability at most
\[
 \sum_{\ell\mid q-1}\ell^{-h}\le(\log_2q)2^{-h}.
\]
Thus $h=O(\log\log q+\log(1/\delta))$ suffices with error $\delta$.
Lift these elements and add the principal-unit generators above.
This avoids factoring $q-1$.

\begin{proof}[Effective operations in Lemma~\ref{lem:effective}]
Standard-form algorithms over finite chain rings compute submodule
membership, sums, intersections, and relations \cite{Georgieva2016}.
Their arithmetic is effective in the presentation of Section~\ref{sec:model};
linear equations in additive coordinates may also be solved by Smith
normal form. Splitness requires the full criterion of
Lemma~\ref{lem:splitting}, not just the dimension of the reduction.
For a split free module, extending a reduced basis and lifting produces
a complement and quotient coordinates.

In an adapted basis, a parabolic is generated by admissible elementary
transvections $I+a_rE_{ij}$, where the $a_r$ additively generate $R$,
and diagonal matrices from generators of $R^\times$.
Elementary elimination over the local ring proves generation of each
diagonal $\GL$ block; the lower blocks are generated by the remaining
transvections.
For $V=W\oplus W'$ the group $(X-I)V\subseteq W$ has matrices
$\left(\begin{smallmatrix}A&B\\0&I\end{smallmatrix}\right)$, and the same
generators in the permitted positions apply.
There are polynomially many generators, and basis conjugation is polynomial.
All costs include the total encoding length of the supplied module generators.
\end{proof}

\section{Transported Fourier procedures}\label{sec:transport}
These procedures explain the relation with \cite{DIKQS14}; they are not
used in the proof of Theorem~\ref{thm:main}.
A finite ring is Frobenius if it admits an additive character whose kernel
contains no nonzero ideal \cite{wood1999duality}.

\subsection{Duality and sampling}

\begin{lemma}\label{lem:perfect}
Let $R$ be a finite chain ring. Then $R$ is a Frobenius ring, and there is a character $\chi:(R,+)\to\C^\times$ whose kernel contains no nonzero ideal, computable from the presentation of $R$. The pairing
\[
  \langle A,B\rangle=\chi\bigl(\tr(AB)\bigr),\qquad A,B\in L:=\Mat_n(R),
\]
is a perfect, symmetric, conjugation-invariant pairing of $(L,+)$ with itself. Consequently, for every subgroup $W\le(L,+)$ one has $|W|\,|W^\perp|=|L|$ and $W^{\perp\perp}=W$.
\end{lemma}

\begin{proof}
Every nonzero ideal contains the simple socle $\pi^{t-1}R$.
Choose a character $\chi$ nontrivial on that socle; it kills no nonzero ideal.
It is computable from additive cyclic coordinates: choose a nonzero
coordinate of $\pi^{t-1}$ and use that coordinate character.
If $\chi(\tr(AB))=1$ for all $B$, take $B=cE_{ji}$ to see that
$\chi(cA_{ij})=1$ for every $c$. Hence $(A_{ij})$ lies in the kernel
and $A_{ij}=0$.
Symmetry and conjugation invariance follow from cyclicity of trace.
The annihilator identities are finite abelian duality.
The pairing identification and its inverse are computable by linear
algebra in additive coordinates, in $\poly(n,\log|R|)$ time.
\end{proof}

\begin{lemma}[Coset sampling]\label{lem:sampling}
Let $W\le(L,+)$, let $\emptyset\ne W'\subseteq W$ and $v_0\in L$. Applying
the abelian QFT of $(L,+)$ to $|v_0+W'\rangle = \frac{1}{\sqrt{|W'|}} \sum_{v\in W'}|v_0+v\rangle$ and measuring yields each
$u\in W^\perp$ with probability exactly $\frac{|W'|}{|W|}\cdot
\frac1{|W^\perp|}$, and hence an element of $W^\perp$ with probability
$|W'|/|W|$. Moreover, if $W'$ is a union of cosets of a subgroup $K\le W$, then every measurement outcome lies in $K^\perp$.
\end{lemma}
 
\begin{proof}
The amplitude at $u$ is $\chi(\tr(v_0u))|W'|^{-1/2}|L|^{-1/2}
\sum_{v\in W'}\langle v,u\rangle$. For $u\in W^\perp$ each summand is $1$,
giving squared amplitude $|W'|/|L|$, and $|L|=|W||W^\perp|$ by
Lemma~\ref{lem:perfect}. For the last claim, suppose $W'=\bigcup_{j}(a_j+K)$ is a union of cosets of \(K\le W\). Then $\sum_{v\in W'}\langle v,u\rangle= \sum_j \langle a_j,u\rangle \sum_{k\in K}\langle k,u\rangle.$ If $u\notin K^\perp$, then $k\mapsto\langle k,u\rangle$ is a nontrivial character of $K$, and hence $\sum_{k\in K}\langle k,u\rangle=0$. Therefore the amplitude at $u$ vanishes, so every measurement outcome lies in $K^\perp$.
\end{proof}

Uniform states over $\GL_n(R)$ can be prepared by testing invertibility
of uniform matrices, with success at least $c_\infty$.
Capping this preparation and using approximate abelian transforms
implements the following sampling routines within their allocated error
budgets.

\subsection{Maximal parabolics and complements}

\begin{proposition}[Maximal parabolics; cf.\ {\cite[Prop.~6]{DIKQS14}}]
\label{prop:maxpar}
Let $q\ge3$ and let $\mathcal H=\{G_{\{U\}}:U\le V\text{ split free},\
0<U<V\}$. Then $\hsp(\GL_n(R),\mathcal H)$ is solvable in quantum time
$\poly(n,\log|R|)$.
\end{proposition}

\begin{proof}
Let $H=G_{\{U\}}$, $\rank U=n-d$, and
$W=\{X\in\Mat_n(R):XU\subseteq U\}$.
Then $H=W\cap G$, $|H|/|W|=c_q(d)c_q(n-d)\ge c_\infty^2$, and
\[
 W^\perp=\{X:XV\subseteq U,\ XU=0\}\cong\Hom_R(V/U,U).
\]
A left coset state $|AH\rangle$ is supported in $AW$.
Lemma~\ref{lem:sampling} gives a uniform element
$u=XA^{-1}\in W^\perp A^{-1}$ with probability $|H|/|W|$.
If $\rank U\le d$, the induced map $V/U\to U$ is surjective with
probability at least $c_\infty$, and $uV=XV=U$.
For a right coset state $|HA\rangle$, the corresponding good sample is
$u=A^{-1}X$. If $d\le\rank U$, that induced map is injective with
split image with probability at least $c_\infty$, giving $\ker u=U$.

At each trial run both the left-coset/image and right-coset/kernel
procedures, since the rank is unknown.
Require a candidate $U'$ to be proper, nonzero, and split free, and test
$G_{\{U'\}}\subseteq H$ on generators.
Theorem~\ref{thm:faithful} forces $U'=U$, so accepted candidates are correct.
A constant success probability and $O(\log(1/\epsilon))$ trials suffice.
\end{proof}

\begin{proposition}[Complements; cf.\ {\cite[Prop.~7]{DIKQS14}}]
\label{prop:complement}
Let $q\ge3$, and let $\mathcal H$ be the family of complements $\mathcal H_\tau=\{I+c\tau\lambda:1+c\in R^\times\}$ of $N_\lambda$ in $G_\lambda$, for $\tau\in V$ with $\lambda(\tau)=1$. Then $\rhsp(G_\lambda,\mathcal H)$ is solvable by a quantum algorithm with $O((1-1/q)^{-(n-1)})$ expected repetitions of a $\poly(n,\log|R|)$-time procedure, returning $R\tau$ at full precision.
\end{proposition}

\begin{proof}
Identify $L_\lambda=\{a\lambda:a\in V\}$ with $V$ and use the
perfect coordinate pairing $\chi(\sum_i a_ib_i)$, rather than the
possibly degenerate restriction of the matrix trace pairing.
Put $W=R\tau\lambda$ and
$W'=\{c\tau\lambda:1+c\in R^\times\}$.
For $A=I+a\lambda$ and $\alpha=1+\lambda(a)\in R^\times$,
subtracting $I$ from a right coset $\mathcal H_\tau A$ gives
$(A-I)+W'\alpha\subseteq(A-I)+W$, with density $1-1/q$.
By Lemma~\ref{lem:sampling}, $n-1$ Fourier outcomes all lie uniformly
in $W^\perp$ with probability $(1-1/q)^{n-1}$; conditioned on that,
they generate this free module with probability at least $c_\infty$.
Compute their $R$-span, then its annihilator and the corresponding
candidate line.

Require the candidate to be split free of rank one with
$\lambda$ surjective on it, and normalize its generator $\tau'$ by
$\lambda(\tau')=1$.
Generate $\mathcal H_{\tau'}$ from $R^\times$ and test its generators
against the oracle. All such complements have order $|R^\times|$,
so containment certifies equality. Since $q\ge3$, an admissible
coefficient $c$ is a unit, and equality of the complements determines
the line. This proves the expected repetition bound.
A fixed multiple of $(1-1/q)^{-(n-1)}\log(1/\epsilon)$ trials gives
the corresponding bounded-error version.
\end{proof}

\begin{remark}\label{rem:complement-q2}
The restriction $q\ge3$ is necessary. Over $\Z/4\Z$ with
$\lambda(x,y)=x$, the vectors $(1,0)$ and $(1,2)$ give the same
complement $\{I,\operatorname{diag}(3,1)\}$.
For $t=1,q=2$, every such complement is trivial.
\end{remark}

\begin{proposition}[Branch B: $d=1$]\label{prop:branchB}
Suppose $\rank T=1$ and $q\ge3$, and draw $\lambda$ uniformly among unimodular elements of $V^*$. With probability $>1-1/q$ the draw is transverse, and then $H\cap G_\lambda$ is a complement of $N_\lambda$ in $G_\lambda$ ($G_\lambda\cong N_\lambda\rtimes (H\cap G_\lambda)$), so Proposition~\ref{prop:complement} returns $T$ at full precision. The expected cost is $\poly(n,\log|R|)\cdot(1-1/q)^{-(n-1)}$; bounded error adds a $\poly(\log(1/\epsilon))$ factor.
\end{proposition}

\begin{proof}
For a transverse $\lambda$, choose $\tau\in T$ with $\lambda(\tau)=1$.
Corollary~\ref{cor:transverse} gives
$H\cap G_\lambda=\mathcal H_\tau$, so the preceding procedure applies.
Lemma~\ref{lem:generic} gives the claimed transversality probability.
Cap each trial of the complement procedure and check its candidate line
against the full oracle using Lemma~\ref{lem:check}.
Thus nontransverse draws cannot cause unbounded runs or false acceptance.
\end{proof}

\begin{remark}\label{rem:branchB-range}
The repetition overhead $(1-1/q)^{-(n-1)}$ is polynomial in $n$
when $n/q=O(\log n)$; at fixed $q$ it is exponential.
Theorem~\ref{thm:hms-lift} replaces this shortcut in the main algorithm.
\end{remark}

\section{Number-field quotient presentations}\label{app:numberfield}
We justify the preprocessing in Corollary~\ref{cor:numberfield}.
The supplied maximal-order basis and multiplication table have total
bit length $L_K$, and $d=[K:\Q]$. Once $N$ is factored, the following
steps use randomized classical polynomial time in $L_K,\log N$.

\paragraph{Prime ideals.}
For each $p\mid N$, work in the finite $\F_p$-algebra
$A=\OK/p\OK$ from the supplied multiplication table.
Compute its nilradical, decompose the reduced algebra into residue fields,
and take the inverse images of the component kernels to obtain the
primes above $p$. The standard prime-decomposition algorithms then give
$p\OK=\prod_{\p\mid p}\p^{e_\p}$; see \cite[\S6.2]{Cohen93}.
The randomized step is finite-field polynomial factorization, for which
Las Vegas expected polynomial time suffices.
An order-computation algorithm is unnecessary because the maximal order
is already supplied. Factoring the defining polynomial modulo $p$ is a
valid shortcut when $p$ does not divide the index of its power order;
that condition, rather than absence of ramification, is the relevant one.

\paragraph{Local rings and CRT.}
Compute the ideal lattices $\p^{a_\p}$ with $a_\p=e_\p v_p(N)$.
Hermite and Smith normal forms \cite[\S2.4]{Cohen93} give quotient
coordinates, additive cyclic decompositions, and effective CRT maps.
Multiplication is inherited from the table of $\OK$.
The identity
$N\OK\subseteq\p^{a_\p}$ bounds the quotient lattice indices by $N^d$;
the ideal and normal-form computations have polynomial bit complexity.
An element of $\p\setminus\p^2$, found by lattice arithmetic, maps to a
uniformizer when $a_\p\ge2$. If $a_\p=1$, the quotient is a field
and we take uniformizer zero.
Residue reduction and a section are computed in these coordinates.
Valuation and exact division by powers of the uniformizer are ideal
membership and linear-equation computations.

\paragraph{Units and error bounds.}
Use the principal-unit generators of Appendix~\ref{app:effective},
together with random lifts generating the residue units with the desired
high probability. No factorization of $q-1$ is needed.
Cap each Las Vegas preprocessing run at a constant multiple of a known
polynomial expectation bound and repeat a logarithmic number of times,
returning failure if all runs time out.
Completed runs have correct output; irreducibility and field decompositions
can be checked by finite-field arithmetic.
Split the preprocessing error budget among these timeouts and the
random unit generating sets. This produces the promised effective
presentations with bounded polynomial time and arbitrarily small error.

For $R=\OK/\p^a$, distinguish the ramification index
$e_{\rm ram}=v_\p(p)$, the length $t=a$, and the characteristic exponent
\[
 e_{\rm char}=\lceil a/e_{\rm ram}\rceil.
\]
Here $a=e_{\rm ram}v_p(N)$, so $e_{\rm char}=v_p(N)$.
The length and characteristic exponent agree when $e_{\rm ram}=1$;
the ramification index is a separate parameter.
Finally,
\[
 \sum_{\p\mid N}a_\p\log|\OK/\p|
 =\sum_{p\mid N}v_p(N)\log p\sum_{\p\mid p}e_\p f_\p
 =d\log N.
\]
There are at most $d\log_2N$ factors.
Together with the error allocation in Corollary~\ref{cor:numberfield},
this proves the stated overall complexity.
The dependence is on the actual input length $L_K$, not on degree alone.

\end{document}